\documentclass[preprint,12pt]{elsarticle}

 \usepackage{amsmath,amsfonts,amsthm,amssymb}

 \usepackage[T1]{fontenc}
 \usepackage{url}

\newtheorem{theorem}{Theorem}
\newtheorem{prop}{Proposition}
 \newtheorem{lemma}{Lemma}
 \newtheorem{observation}{Observation}
 \newtheorem{definition}{Definition}
 \newtheorem{claim}{Claim}
 \newtheorem{corollary}{Corollary}

 \usepackage{algorithm}
\usepackage{algpseudocodex}

\usepackage{newunicodechar}
\newunicodechar{≤}{\ensuremath{\leq}}
\usepackage{framed}

\newcommand{\decisionproblem}[3]{%
\begin{framed}
\vspace{0,1cm} \noindent
 \textbf{\textsc{#1} }
\par%
\noindent
{\bf{Input:}}  #2\\%
{\bf{Output:}} #3%
\end{framed}
}
\newcommand{\decisionproblemparam}[4]{%
\begin{framed}
\vspace{0,1cm} \noindent 
 \textbf{\textsc{#1} }
  \vspace{1.2mm}%
\par%
\noindent
{\bf{Input:}}  #2\\%
{\bf{Parameter:}}  #3\\%
{\bf{Output:}} #4%
\end{framed}
}

 \usepackage{lineno}
\journal{Journal of Computer and System Sciences}

\begin{document}

\begin{frontmatter}



\title{Structural Parameters for Dense Temporal Graphs}


\author[label1]{Jessica Enright}
\author[label1]{Samuel D. Hand}
\author[label1]{Laura Larios-Jones} 
\author[label1]{Kitty Meeks}

\affiliation[label1]{organization={School of Computing Science, University of Glasgow},
            addressline={Sir Alwyn Williams Building}, 
            city={Glasgow},
            postcode={G12 8RZ}, 
            state={Scotland},
            country={UK}}

\begin{abstract}
Temporal graphs provide a useful model for many real-world networks. Unfortunately the majority of algorithmic problems we might consider on such graphs are intractable.  There has been recent progress in defining structural parameters which describe tractable cases by simultaneously restricting the underlying structure and the times at which edges appear in the graph.  These all rely on the temporal graph being sparse in some sense.  We introduce temporal analogues of three increasingly restrictive static graph parameters -- cliquewidth, modular-width and neighbourhood diversity -- which take small values for highly structured temporal graphs, even if a large number of edges are active at each timestep.  The computational problems solvable efficiently when the temporal cliquewidth of the input graph is bounded form a subset of those solvable efficiently when the temporal modular-width is bounded, which is in turn a subset of problems efficiently solvable when the temporal neighbourhood diversity is bounded.  By considering specific temporal graph problems, we demonstrate that (up to standard complexity theoretic assumptions) these inclusions are strict.
\end{abstract}




\end{frontmatter}


\section{Introduction}

Temporal graphs, in which the set of active edges changes over time, are a useful formalism for modelling numerous real-world phenomena, from social networks in which friendships evolve over time to transport networks in which a particular connection is only available on particular days and times.  This has inspired a large volume of research into the algorithmic aspects of such graphs in recent years \cite{casteigts_temporal_2021,holme_temporal_2012,michail_introduction_2016}, but unfortunately in many cases even problems which admit polynomial-time algorithms on static graphs become intractable in the temporal setting.  

This has motivated the study of computational problems on restricted classes of temporal graphs, with mixed success: in a few cases, restricting the structure of the underlying static graph (e.g. to be a path or a tree) is effective, but numerous natural temporal problems remain intractable even when the underlying graph is very strongly restricted (e.g. when it is required to be a path \cite{mertzios_computing_2023} or a star \cite{akrida_temporal_2021}).  Recently, several promising new parameters have been introduced that simultaneously restrict properties of the static underlying graph and the times at which edges are active in the graph; these include several analogues of treewidth for temporal graphs \cite{fluschnik_as_2020,mans_treewidth_2014}, the temporal feedback edge/connection number \cite{haag_feedback_2022}, the timed vertex feedback number \cite{casteigts_finding_2021} and the (vertex-)interval-membership-width of the temporal graph \cite{bumpus_edge_2023}.  
However, all of these new temporal parameters are only small for temporal graphs that are, in some sense, sparse: none of them can be bounded on a temporal graph which has a superlinear (in the number of vertices) number of active edges at every timestep.  

In this paper, we attempt to fill this gap in the toolbox of parameters for temporal graphs by introducing three new parameters which can take small values on temporal graphs which are dense but are sufficiently highly structured.  Specifically, we define natural temporal analogues of cliquewidth, modular-width and neighbourhood diversity, all of which have proved highly effective in the design of efficient algorithms for static graphs.  

Importantly, the neighbourhood diversity of a static graph upper bounds its modular-width, which upper bounds its cliquewidth.  Both cliquewidth (introduced by Courcelle et al. \cite{courcelle_handle-rewriting_1993}) and modular-width (introduced by Gajarsk\'{y}~et~al.~\cite{gajarsky_parameterized_2013}, using the long-standing notion of modular decompositions \cite{gallai_transitiv_1967}) can be defined in terms of width measures over composition trees allowing particular operations.  Cliquewidth constructions have greater flexibility due to the fact we are allowed to use an additional ``relabelling'' operation; this makes it possible, for example, to build long induced paths, which cannot exist in graphs of small modular width.  Courcelle et al. \cite{courcelle_linear_2000} show that any graph property expressible in monadic second order is solvable in linear time for graphs of bounded cliquewidth. Gajarsk\'{y}~et~al.~\cite{gajarsky_parameterized_2013} provide examples of problems (\textsc{Hamilton Path} and \textsc{Colouring}) that are hard with respect to cliquewidth but tractable with respect to the more restrictive parameter modular-width.  Neighbourhood diversity is a highly restrictive parameter, introduced by Lampis \cite{lampis_algorithmic_2012}, which requires that large sets of vertices have identical neighbourhoods.  Cordasco \cite{cordasco_iterated_2020} demonstrated that \textsc{Equitable Colouring} is hard with respect to modular-width but tractable with respect to neighbourhood diversity.

These three static parameters are the inspiration for our temporal parameters. 
 Informally, our new parameters are defined as follows:
 
\begin{itemize}
    \item A temporal graph has \emph{temporal neighbourhood diversity (TND)} at most $k$ if its vertices can be partitioned into at most $k$ classes such that each class induces either an independent set or a clique in which all edges are active at exactly the same times, and any two vertices in the same class have exactly the same neighbours outside the class at each timestep.
    \item \emph{Temporal modular-width (TMW)} is a generalisation of TND: a temporal graph has TMW at most $k$ if its vertices can be partitioned into modules such that two vertices in the same module must have exactly the same neighbours outside the class at each timestep, but now each module need only be a temporal graph which itself has TMW at most $k$, rather than a clique or independent set.
    \item Like the static version, \emph{temporal cliquewidth (TCW)} is defined to be the minimum number of labels needed to construct a temporal graph using four operations (create a vertex with a new label; take a disjoint union of two graphs; add all edges between vertices of two specified labels; relabel all occurrences of one label to another); the difference from the static case is that when adding edges between two sets of vertices these edges must all be active at exactly the same times.
\end{itemize}

We note that in every case we will recover the corresponding static parameter if all edges are active at the same times.  It is immediate that the TND of a temporal graph is an upper bound on its TMW, and it is straightforward to show (see Section \ref{sec:TMW}) that the TMW is an upper bound on the TCW.  Thus, the most general algorithmic result we hope to obtain is to show that a problem is tractable when the TCW of the input temporal graph is bounded, but we expect to be able to show tractability for more problems as we impose stronger restrictions on the input by bounding respectively the TMW and the TND.  

\begin{figure}[ht]
    \centering
    \includegraphics[page=1, width =0.8\textwidth]{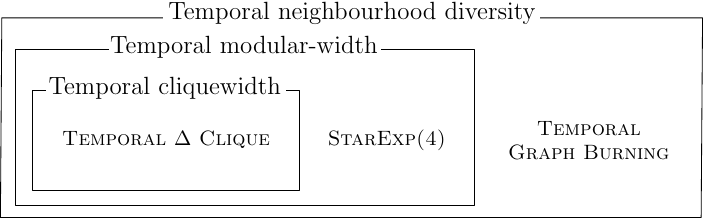}
    \caption{A diagram of our parameters and the problems we show to be tractable with respect to each. A problem is in a rectangle if it is tractable with respect to the parameter it is labelled with. Assuming P$\neq$NP, each problem is in the rectangle for the most general of the three parameters for which it is tractable.}
    \label{fig:target}
\end{figure}

To illustrate the value of considering this hierarchy of parameters, we provide examples of problems which can be solved efficiently when each of our three new parameters is bounded, but which (in the case of TND and TMW) remain intractable when we restrict only the next most restrictive parameter, as illustrated in Figure \ref{fig:target}.  Specifically, we prove that:
\begin{itemize}
    \item \textsc{Temporal $\Delta$ Clique} is solvable in linear time when a temporal cliquewidth decomposition of constant width is given (see Section \ref{sec:TCW}).
    \item \textsc{StarExp(4)} (the problem of deciding whether there is a closed temporal walk visiting all vertices in a star when each edge is active at no more than four times) remains NP-hard on graphs with TCW at most three, but is in FPT with respect to the TMW of the input graph\footnote{We use fpt (lowercase) as a descriptor of algorithms witnessing the inclusion of a problem in the parameterised complexity class FPT.} (see Section \ref{sec:TMW}).
   \item \textsc{Graph Burning} is NP-hard on temporal graphs with constant TMW, but is in FPT with respect to the TND of the input graph (see Section \ref{sec:TND}).
\end{itemize}
We also (in Section \ref{sec:TND}) provide an fpt-algorithm to solve \textsc{SingMinReachDelete} parameterised by TND and the maximum number of appearances of any edge, in order to illustrate additional techniques that may be used when working with this new temporal parameter. This problem asks, for a given source vertex, what the cardinality of a smallest set of time-edges is such that their deletion leaves at most $r$ vertices temporally reachable from the source.  We conjecture that \textsc{SingMinReachDelete} is another example of a problem that is tractable with respect to TND but intractable when only the TMW is restricted.

The remainder of the paper is organised as follows.  We conclude this section by introducing some key notation and definitions used throughout the paper.  The following three sections are devoted to TCW, TMW and TND respectively, with each section containing the formal definition of a parameter as well as results about problems which can be solved efficiently when that parameter is bounded.  

\subsection{Notation and definitions}

We use a number of standard notations for temporal graphs and related notions.  A \emph{temporal graph} $\mathcal{G} = (G, \lambda)$ consists of an underlying static graph $\mathcal{G}_\downarrow=G$, and a time-labelling function $\lambda : E \to 2^\mathbb{N}$, assigning to each edge a set of timesteps at which it is active. We refer to a pair $(e,t)$ consisting of an edge $e\in E(G)$ and time $t\in\lambda(e)$ as a \emph{time-edge}. The set of all time-edges of a temporal graph is denoted by $\varepsilon(\mathcal{G})$ and the \emph{lifetime} $\Lambda$ of a temporal graph refers to the final time at which any edge is active, i.e.  $\Lambda = \max \{ \max \lambda(e) : e \in  E(G)\}$. The \emph{snapshot} $\mathcal{G}_t$ of a temporal graph $\mathcal{G}$ at time $t$ is the static graph $G=(V,E_t)$ where $E_t$ is the set of edges active at time $t$.

A \emph{temporal path} on the temporal graph $\mathcal{G}=(G, \lambda)$ is a sequence of edge, time pairs $(e_1,t_1),...,(e_\ell,t_\ell)$, such that $e_1,...,e_\ell$ is a path on $G$, and $t_i\in \lambda(e_i)$ for every $i\in[\ell]$, and $t_1,...,t_\ell$ is a strictly increasing sequence of times.  Given a temporal path $(e_1,t_1),...,(e_\ell,t_\ell)$ we refer to the time $t_1$ as its \emph{departure time}, and $t_\ell$ as its \emph{arrival time}.

We refer the reader to \cite{cygan_parameterized_2015} for standard definitions from the field of parameterised complexity.

\section{Tractability with respect to Temporal Cliquewidth}\label{sec:TCW}

In this section we give the formal definition of the first of our new parameters, temporal cliquewidth. We later demonstrate that the problem of finding a temporal clique admits an fpt-algorithm parameterised by temporal cliquewidth.  Before defining temporal cliquewidth, we start by recalling the definition of cliquewidth in the static setting, as introduced by Courcelle and Olariu \cite{courcelle_upper_2000}. 

\begin{definition}[Cliquewidth]\label{def:cw}
The \emph{cliquewidth} of a static graph $G=(V,E)$ is the number of labels required to construct $G$ using only the following operations:
    \begin{enumerate}
    \item Creating a new vertex with label $i$.
    \item Taking the disjoint union of two labelled graphs.
    \item Adding edges to join all vertices labelled $i$ to all vertices labelled $j$, where $i \neq j$.
    \item Renaming label $i$ to label $j$.
\end{enumerate}
We refer to an algorithm which constructs a graph $G$ using the above operations as a \emph{cliquewidth construction} of $G$.
\end{definition}

Computing the cliquewidth of a graph is NP-hard \cite{fellows_clique-width_2009}, although there exists a polynomial-time algorithm to recognise graphs of cliquewidth at most three \cite{corneil_polynomial-time_2012}.

Translating this definition into the temporal setting, and preserving the idea that vertices with the same label should be indistinguishable when we add edges -- which imposes additional restrictions on the times at which new edges are active -- we obtain our definition of temporal cliquewidth.
\begin{definition}[Temporal Cliquewidth]\label{def:temp-cw}
The \emph{temporal cliquewidth} of a temporal graph $\mathcal{G}=(G, \lambda)$ is the number of labels required to construct $\mathcal{G}$ using only the following operations:
    \begin{enumerate}
    \item Creating a new vertex with label $i$.
    \item\label{cliquedisjoint} Taking the disjoint union of two labelled graphs.
    \item Adding edges to join all vertices labelled $i$ to all vertices labelled $j$, where $i \neq j$, such that all the added edges are active at the same set of times $T$.
    \item Renaming label $i$ to label $j$.
\end{enumerate}
We refer to an algorithm which constructs a temporal graph $\mathcal{G}$ using the above operations as a \emph{temporal cliquewidth construction} of $\mathcal{G}$.
\end{definition}
We note that, if temporal graph $\mathcal{G}$ has bounded temporal cliquewidth $k$, then the underlying graph $\mathcal{G}_{\downarrow}$ of $\mathcal{G}$ has cliquewidth at most $k$. The construction of the underlying graph is found by adding a static edge (if one does not already exist) whenever a time-edge is added in the construction of the temporal graph. In addition, the snapshot of $\mathcal{G}$ at any time $t$ also has cliquewidth at most $k$. Furthermore, given a temporal graph $\mathcal{G}$ where all edges are active at the same set of times, the temporal cliquewidth of $\mathcal{G}$ must be exactly the cliquewidth of the underlying graph $\mathcal{G}_{\downarrow}$. It follows immediately that it is NP-hard to compute temporal cliquewidth, as the NP-hard problem of computing the cliquewidth of a static graph \cite{fellows_clique-width_2009} is a special case. 

A cliquewidth construction tree can be built by adding a node for each operation, starting with a leaf for each new vertex added. This is a rooted tree, where the root node corresponds to the entire constructed graph. For some node $s$, we use ``parent'' to refer to the unique node adjacent to $s$ which is on the path from $s$ to the root. A node $s_p$ is the parent of another node $s_c$ if the node $s_p$ corresponds to the outcome of performing operation (2), (3) or (4) on $s_c$.  We note that the number of nodes in the temporal cliquewidth construction tree is $O(nk^2)$. To see this, note that there is exactly one leaf node for each introduced vertex. In addition, the nodes corresponding to the disjoint union operation are the only nodes that can have degree greater than two. This means that any path consisting of nodes with degree exactly two must consist of nodes corresponding to only relabelling and adding time edges. This bounds the length of such paths by $O(k^2)$.

We begin by showing that \textsc{Temporal $\Delta$ Clique} is in FPT parameterised by the temporal cliquewidth of the input graph. The latter problem was introduced by Viard et al. \cite{viard_computing_2016} and asks, in an interval of times, whether there is a set of at least $h$ vertices such that there is an appearance of an edge between every pair of vertices in every sub-interval of $\Delta$ consecutive times. Note that, if the interval of interest is not the entire lifetime of the temporal graph, one can remove the appearances of edges which fall outside of this interval in time linear in the number of edges. Hermelin et al. \cite{hermelin_temporal_2023} investigate the variant of this problem where the interval in question is the entire lifetime of the temporal graph. More formally, they ask if there exists a set $V'$ of cardinality at least $h$ such that for each pair of distinct vertices $u,v\in V'$ and each time $0\leq i\leq \Lambda-\Delta+1$ where $\Lambda$ is the lifetime of $\mathcal{G}$, there exists a time-edge at time $t'\in[i,\Delta+i-1]$.

Hermelin et al. note that this is the case if and only if there is a set of vertices of size at least $k$ such that they form a clique on the static graph consisting of edges that appear in every interval of $\Delta$ timesteps. They name this static graph a $\Delta$\emph{-association graph}. It is more formally defined as $G=\left(V,\,\bigcap^{\Lambda(\mathcal{G})-\Delta+1}_{i=1}\bigcup^{i+\Delta-1}_{j=i}E_j\right)$ where $E_j$ is the set of edges active at time $j$. This reformulates the problem as follows.

\newpage

\decisionproblem{Temporal $\Delta$ Clique}{A temporal graph $\mathcal{G}=(V,E,\lambda)$ and two integers $\Delta$ and $h$ where $\Delta\leq T(\mathcal{G})$.}{Is there a set $V'\subseteq V$ of vertices such that $|V'|\geq h$ and $V'$ is a clique in the $\Delta$-association graph $G$ of $\mathcal{G}$?}

Note that a temporal graph $\mathcal{G}$ with integers $\Delta$ and $h$ is a yes-instance of \textsc{Temporal $\Delta$ Clique} if and only if its $\Delta$-association graph $G$ and $h$ are a yes-instance of \textsc{Clique}.
\begin{prop}
     If a temporal graph $\mathcal{G}$ has temporal cliquewidth $k$, then the $\Delta$-association graph $G$ of $\mathcal{G}$ has cliquewidth at most $k$.
\end{prop}
\begin{proof}
    Given a temporal cliquewidth construction of $\mathcal{G}$, we find a cliquewidth construction of its $\Delta$-association graph $G$ using the same number of labels. We do so by modifying the temporal cliquewidth construction of $\mathcal{G}$. Call this construction $\Omega$. We will denote by $\Omega_G$ the cliquewidth construction of $G$. All instances of operations 1, 2, and 4 in $\Omega$ are kept the same in $\Omega_G$. For every operation that adds a time-edge $(e,t)$ such that $e\in E(G)$ and no earlier operation in $\Omega$ adds a time-edge containing $e$, we add an operation which adds edges between the same sets of vertices to $\Omega_G$. That is, edges are added to $G$ such that, for each edge $e$ in $G$, $e$ is added in $\Omega_G$ the first time a time-edge containing $e$ is added in the construction $\Omega$. 

    We first show that there are no edges added by $\Omega_G$ which are not in $E(G)$. Later, we show that all edges in $E(G)$ are added by $\Omega_G$.

     We note that edges are added to $G$ by $\Omega_G$ only if there is an operation in $\Omega$ that adds a set of time-edges containing at least one time-edge $(e,t)$ such that $e$ is in $E(G)$ and has not yet been added to the graph. Suppose, for a contradiction, that there exists a time-edge $(e',t')$ which is added to $\mathcal{G}$ by this operation such that $e'\notin E(G)$. Since this operation is the first time a time-edge containing $e$ is added, the times at which $e$ is active form a subset of the times at which $e'$ is active. We know that $e$ appears at least once in each interval of size $\Delta$ by definition of $G$. Thus, so must $e'$. Hence, $e'$ is in $E(G)$; a contradiction.

     It is clear from our construction of $\Omega_G$ that an edge is added if it is in the graph $G$. For all edges in $G$, there is at least one time-edge in $\mathcal{G}$ containing them. Thus, the operation adding the edges is added to $\Omega_G$ when time-edges containing the edges are first added in $\Omega$. Thus, all edges in $E(G)$ are added in the construction $\Omega_G$; $\Omega_G$ constructs $G$. 
     
    This proves that, given a temporal cliquewidth construction of $\mathcal{G}$, we find a cliquewidth construction of its $\Delta$-association graph $G$ using the same number of labels. 
\end{proof}
Gurski \cite[Theorem 4.4]{gurski_comparison_2008} shows that \textsc{Clique} is solvable in linear time in graphs of bounded cliquewidth if its construction is given. This gives us the following result.
\begin{theorem}
    Given a cliquewidth construction of the $\Delta$-association graph of a temporal graph $\mathcal{G}$, \textsc{Temporal $\Delta$ Clique} can be solved in linear time.
\end{theorem}

\section{Tractability with respect to Temporal Modular-width}\label{sec:TMW}

We now introduce a more restrictive parameter, temporal modular-width, and show (in Section \ref{sec:starexp}) that there exist problems which are efficiently solvable when this parameter is bounded even though they remain intractable on temporal graphs with constant temporal cliquewidth.

We begin with the formal definition of the parameter.  Again, we start by recalling the definition of the corresponding static parameter on which our definition is based. 
\begin{definition}[Modular-width, Section 2.5 \cite{gajarsky_parameterized_2013}]\label{def:mw}
    Suppose a static graph $G$ can be constructed by the algebraic expression $A$ which uses the following operations:
    \begin{enumerate}
        \item Creating an isolated vertex.
        \item\label{modular-disjoint} Taking the disjoint union of two graphs.
        \item\label{modular-complete} Taking the complete join of two graphs. That is, for graphs $G_1=(V_1,E_1)$ and $G_2=(V_2,E_2)$, $V(G_1\otimes G_2)=V(G_1)\cup V(G_2)$ and $E(G_1\otimes G_2)=E(G_1)\cup E(G_2)\cup \{(v,w):v\in V(G_1) \text{ and }w\in V(G_2)\}$.
        \item\label{step4static} The substitution of graphs $G_1,\ldots, G_n$ into a graph $G'$ with vertices $v_1,\ldots,v_n$. This gives the graph $G'(G_1,\ldots,G_n)$ with vertex set $\bigcup_{1\leq i\leq n}V(G_i)$ and edge set $\bigcup_{1\leq i\leq n}E(G_i)\cup \{(v,w):v\in V(G_i), w\in V(G_j),\text{ and }(v_i,v_j)\in E(G')\}$.
    \end{enumerate}
    The width of an expression $A$ is the maximum number of operands in an occurrence of the operation \ref{step4static} in $A$. The \emph{modular-width} of $G$, written $MW(G)$, is this minimum width of an expression $A$ which constructs $G$.
\end{definition}
We refer to the graphs $G_1,\ldots, G_n$ which we substitute into $G'$ as \emph{modules}.  It is known that for any graph $G$ an algebraic expression of modular-width $MW(G)$ can be found in linear time \cite{mcconnell_modular_1999,tedder_simpler_2008}. Observe that operations \ref{modular-disjoint} and \ref{modular-complete} are special cases of operation \ref{step4static}.

We now define our temporal analogue of this parameter.

\begin{definition}[Temporal Modular-width]\label{def:temp-mw}
    Suppose a temporal graph $\mathcal{G}$ can be constructed by the algebraic expression $A$ which uses the following operations:
    \begin{enumerate}
        \item\label{step1temp} Creating an isolated vertex.
         \item\label{step2temp} Taking the disjoint union of two temporal graphs.
         \item\label{step3temp} Taking the complete join of two temporal graphs at a set of times $T$. That is, for graphs $\mathcal{G}_1=((V_1,E_1),\lambda_1)$ and $\mathcal{G}_2=((V_2,E_2),\lambda_2)$, $V(\mathcal{G}_1\otimes \mathcal{G}_2)=V(\mathcal{G}_1)\cup V(\mathcal{G}_2)$ and $\varepsilon(\mathcal{G}_1\otimes \mathcal{G}_2)=\varepsilon(\mathcal{G}_1)\cup \varepsilon(\mathcal{G}_2)\cup \{(vw,t):v\in V(\mathcal{G}_1), w\in V(\mathcal{G}_2)\text{ and } t\in T\}$.
        \item\label{step4temp} The substitution of temporal graphs $\mathcal{G}_1,\ldots, \mathcal{G}_n$ into a temporal graph $\mathcal{G}'$ with vertices $v_1,\ldots,v_n$. This gives the graph $\mathcal{G}'(\mathcal{G}_1,\ldots,\mathcal{G}_n)$ with vertex set $\bigcup_{1\leq i\leq n}V(\mathcal{G}_i)$ and time-edge set $\bigcup_{1\leq i\leq n}\varepsilon(\mathcal{G}_i)\cup \{(vw,t):v\in V(\mathcal{G}_i), w\in V(\mathcal{G}_j),\text{ and }(v_iv_j,t)\in \varepsilon(\mathcal{G}')\}$.
    \end{enumerate}
    The width of an expression $A$ is the maximum number of operands in an occurrence of the operation \ref{step4temp} in $A$. The \emph{temporal modular-width} of $(G,\lambda)$ is this minimum width of an expression $A$ which constructs $(G,\lambda)$.
\end{definition}

As for temporal cliquewidth, we observe that the temporal modular-width of a temporal graph is equal to the modular-width of the underlying graph if all edges have the same temporal assignment.  It follows that, as in the static case, the temporal modular-width bounds the length of the longest induced path in the underlying graph. To see this, note that we can only construct a long path by substitution into a path, otherwise we would have a disconnected graph or introduce edges between vertices which are not adjacent in the path.

We define a modular-width construction tree analogously to the cliquewidth construction tree used earlier.

We now argue that, as claimed, bounding the temporal modular-width of a temporal graph is a strictly stronger restriction than bounding the temporal cliquewidth.

\begin{theorem}
    For any temporal graph $\mathcal{G} = (G, \lambda)$, the temporal cliquewidth is upper bounded by the temporal modular-width.
\end{theorem}
\begin{proof}
    Assuming that $\mathcal{G}$ has temporal modular-width at most $k$, we induct on the temporal modular decomposition tree for $\mathcal{G}$ to produce a temporal cliquewidth construction of $\mathcal{G}$ using at most $k$ labels. This is achieved by inductively producing a temporal cliquewidth construction corresponding to each child of the root node in the temporal modular decomposition tree, and then relabelling and combining these constructions to produce an overall expression for $\mathcal{G}$.
    
    Any tree of depth one consists of only a single vertex and thus has temporal cliquewidth $1 \leq k$.

    Now, given a tree of depth $d$, consider the substitution $\mathcal{G'}(\mathcal{G}_1, ..., \mathcal{G}_n)$ at the root node of the tree. We can make this assumption without loss of generality by recalling that operations 2 and 3 are special cases of substitution into 2-vertex graphs. Note that $n \leq k$ as $\mathcal{G}$ has temporal modular-width at most $k$. Then each child $\mathcal{G}_i$ has a decomposition tree of depth at most $d-1$, and therefore by induction has cliquewidth at most $k$. Therefore it is possible to find an expression $A_i$ for every child $\mathcal{G}_i$ using the operations of Definition~\ref{def:temp-cw}, such that at most $k$ labels are used. 
    The graph $\mathcal{G}$ can then be constructed by relabelling every vertex in each $A_i$ to $i$, and then taking a disjoint union, before adding edges active at times $\lambda(\{v_i,v_j\})$ between all vertices labelled $i$ and all vertices labelled $j$ for every $\{v_i,v_j\}\in E(G)$, where $\mathcal{G}'=(G, \lambda)$.
\end{proof}

We now show that the unique decomposition of the graph into maximally sized temporal modules, and therefore the width, can be computed in polynomial time. Habib and Paul \cite{DBLP:journals/csr/HabibP10} describe a simple algorithm for finding the modular decomposition of a static graph. This operates by finding and repeatedly adding \textit{splitters} to a candidate module. We use a similar method to find the temporal modular decomposition. We begin by recalling the definition of splitters, and use this to give a definition for temporal modules.

\begin{definition}[Splitter (Habib and Paul \cite{DBLP:journals/csr/HabibP10})]
    Given a set of vertices $S$, a vertex $x$ is said to be a \emph{splitter} for $S$ if there exist vertices $u, v \in S$ such that $x$ is adjacent to exactly one of $u$ and $v$.
\end{definition}

\begin{definition}[Static Module (Habib and Paul \cite{DBLP:journals/csr/HabibP10})]
    A set of vertices $M$ is a \emph{module} of the static graph $G$, if and only if for every vertex $x \notin M$, $x$ is not a splitter for $M$.
\end{definition}

\begin{definition}[Temporal Module]
    A set of vertices $M$ is a \emph{temporal module} of $(G, \lambda)$ if and only if for every timestep $t \in [\Lambda]$ and vertex $x \notin M$, $x$ is not a splitter for $M$ in the snapshot graph $(V(G), \{e \in E(G) : t \in \lambda(e) \})$.
\end{definition}

We now show that every temporal module of $\mathcal{G}$ is a module of every static snapshot graph of $\mathcal{G}$, and conversely, that any vertex set which is a module of every snapshot graph is a temporal module of $\mathcal{G}$.

\begin{lemma}
    \label{lem:modsnap1}
    $M$ is a temporal module of $(G, \lambda)$ if and only if for every timestep $t \in [\Lambda]$, $M$ is a module of the snapshot graph $(V(G), \{e \in E(G) : t \in \lambda(e) \})$.
\end{lemma}
\begin{proof}
    If $M$ is a temporal module of $(G, \lambda)$, then for any timestep $t$ there are no splitters of $M$ in $(V(G), \{e \in E(G) : t \in \lambda(e) \})$, and thus $M$ is a module of $(V(G), \{e \in E(G) : t \in \lambda(e) \})$.

    Conversely, if for every timestep $t$, $M$ is a module of $(V(G), \{e \in E(G) : t \in \lambda(e) \})$, it must have no splitters in $(V(G), \{e \in E(G) : t \in \lambda(e) \})$, and thus is a temporal module of $(G, \lambda)$.
\end{proof}

We further demonstrate that, as in the static case, the set of maximal modules for a temporal graph is unique.

\begin{lemma}
    The set of maximal temporal modules $\mathcal{M}$ for the temporal graph $(G, \lambda)$ is unique and partitions $V(G)$.
\end{lemma}
\begin{proof}
    Assume that such a set does not partition $V(G)$, that is $\bigcup \mathcal{M} = V(G)$ but there exists $M_i, M_j \in \mathcal{M}$ such that $M_i \cap M_j \neq \emptyset$. By the previous lemma, we have that $M_i$ and $M_j$ are modules in every snapshot of $(G, \lambda)$, and hence $M_i \cup M_j$ is also a module in every snapshot of $(G, \lambda)$. This follow from the fact that there are no vertices that are a splitter for either $M_i$ or $M_j$ in any snapshot, so no vertex can split their union in any snapshot. This means that $M_i \cup M_j$ is therefore a temporal module, but $M_i \cup M_j \supset M_i$, contradicting the maximality.

    Now assume that such a set is not unique, that is there exists a set of maximal temporal modules $\mathcal{P}$ with $\mathcal{P} \neq \mathcal{M}$. Now let $P_i \in \mathcal{P}$ be a module such that $P_i \notin \mathcal{M}$, and consider a vertex $v \in P_i$. As $\mathcal{M}$ partitions $V(G)$, there exists some module $M_j \in \mathcal{M}$ such that $v \in M_j$. Then $P_i \cup M_j$ is a module in every snapshot, and therefore a temporal module, but this contradicts the maximality of $\mathcal{M}$ and $\mathcal{P}$.
\end{proof}

A simple way to compute the unique maximal modular decomposition is then provided by the following observation.

\begin{observation}
    Let $S$ be a subset of the vertices of a temporal graph $(G, \lambda)$. If $S$ has a splitter $x$ in any snapshot of $(G, \lambda)$, then any module of $(G, \lambda)$ containing $S$ also contains $x$.
\end{observation}

We now recursively compute the modular decomposition as follows. Given a set $\mathcal{M}$ of modules, repeatedly test if a non-trivial module containing a pair $M_1, M_2 \in \mathcal{M}$ exists, by considering $M_1 \cup M_2$ as a candidate module, and repeatedly adding any remaining modules in $\mathcal{M}$ containing a vertex that splits $M_1 \cup M_2$. If a trivial module is obtained the pair is discarded, and otherwise the set of modules is updated with the newly found non-trivial module. We find the maximal modular decomposition by initialising the set of modules to all the singleton trivial modules, that is $\{ \{v\} : v \in V(G) \}$.

\begin{theorem}
    We can find a temporal modular decomposition of minimum width in time $O(n^4\Lambda)$, where $n$ is the number of vertices in the temporal graph.
\end{theorem}
\begin{proof}
    The above algorithm begins with a set of $n$ modules, and repeatedly considers pairs from this set. After each iteration the size of the list will either be reduced by one, or a pair will be discounted, thus the process must terminate after $O(n^2)$ iterations. On each iteration each remaining module is checked to see if it splits the current candidate module on any timestep, which can be done in time $O((n+m)\Lambda) = O(n^2\Lambda)$, giving an overall runtime of $O(n^4\Lambda)$.
\end{proof}

\subsection{Star Exploration}\label{sec:starexp}

In this section we consider the following problem, demonstrating that it remains NP-hard even on temporal graphs with temporal cliquewidth at most three, but that it is solvable in constant time on graphs with bounded temporal modular-width (provided we are given the temporal modular-width of the graph). This problem was first introduced by Akrida et al. \cite{akrida_temporal_2021}.

\decisionproblem{StarExp($\tau$)}{A temporal star $(S_n, \lambda)$ where $| \lambda(e) | \leq \tau$ for every edge $e$ in the star $S_n$.}{Is there a strict temporal walk, starting and finishing at the centre of the star, which visits every vertex of $S_n$?}

We begin with a simple observation about the temporal cliquewidth of temporal graphs whose underlying graph is a star.

\begin{lemma}\label{lem:star-cw}
    A temporal star $S_n$ has temporal cliquewidth at most 3.
\end{lemma}
\begin{proof}
    Our proof is constructive. The temporal cliquewidth construction is as follows.
    \begin{enumerate}
        \item Order the leaves of the star $l_1,\ldots,l_n$ arbitrarily.
        \item Introduce the central vertex $c$ with label $L_1$.
        \item Introduce the vertex $l_1$ with label $L_2$ and add all time-edges between $l_1$ and $c$.
        \item For all remaining leaves $l_i \in l_2,\ldots,l_n$:
        \begin{enumerate}
            \item Introduce $l_i$ with label $L_3$ and all time-edges between $c$ and $l_i$.
            \item Relabel $l_i$ with label $L_2$.
        \end{enumerate}
    \end{enumerate}
    From this construction, we see that it requires at most 3 labels to construct a temporal star.
\end{proof}
\textsc{StarExp($\tau$)} is known to be NP-hard even for constant $\tau$ \cite{akrida_temporal_2021, bumpus_edge_2023}. Then, by Lemma \ref{lem:star-cw}, \textsc{StarExp($\tau$)} is an example of a problem which is NP-hard on graphs of bounded temporal cliquewidth. We now show that \textsc{StarExp($\tau$)} is tractable on graphs of bounded temporal modular-width. We begin with the following lemma.

\begin{lemma}\label{lem:star-group-leaves}
    If a temporal star $S_n$ has temporal modular-width at most $k$, the leaves of $S_n$ can be partitioned into $k-1$ subsets such that, if $u$ and $v$  are in the same subset and $c$ is the central vertex in the star, the edges $uc$ and $cv$ are active at the same times.
\end{lemma}
\begin{proof}
    We claim that, when constructing a star, the graph into which the final substitution is made in the maximal temporal modular decomposition is a star with at most $k-1$ leaves. Else, the graph constructed would either not be a star or not have temporal modular-width at most $k$. We begin by showing that, for this substitution, the central vertex $c$ is the only vertex in its module. Suppose otherwise; following the substitution any other vertex in this module has a neighbour which is not $c$. This contradicts that $S_n$ is a star.

    In addition, we have that the other modules are independent sets, otherwise we would again have that two leaves are adjacent. Therefore, there are no edges in the graph before this final substitution is made. Since all edges between any two modules in a substitution are assigned the same time, all leaves in the same module must be adjacent to the central vertex by edges active at the same time. Thus, we have $k-1$ subsets such that, if $u$ and $v$  are in the same subset, the edges $uc$ and $cv$ are active at the same times.
\end{proof}
\begin{lemma}\label{lem:star-no}
    If there are strictly more than $\tau/2$ leaves of $S_n$ whose incident edges are active at the same times, we have a no-instance of \textsc{StarExp($\tau$)}.
\end{lemma}
\begin{proof}
    By definition of the problem \textsc{StarExp($\tau$)}, each edge in the star is active at most $\tau$ times. Therefore, if there are $\lfloor \tau/2 \rfloor + 1$ vertices $u_1, \dots, u_{\lfloor \tau/2 \rfloor + 1}$ such that $\lambda(u_1c)= \dots =\lambda(u_{\lfloor \tau/2 \rfloor + 1}c)$ where $c$ is the central vertex in the star, there is no temporal walk which starts at $c$ and visits all of $u_1, \dots, u_{\lfloor \tau/2 \rfloor + 1}$. To see this, note that visiting a leaf and returning requires the use of two distinct time-edges. Therefore, a walk visiting any $\lfloor \tau/2 \rfloor + 1$ leaves which departs from and returns to $c$ must consist of at least $\tau + 1$ distinct time-edges. This is not possible if these vertices have incident edges that are active at the same times and $|\lambda(uc)|\leq \tau$.
\end{proof}
\begin{theorem}\label{thm:star-tractable-modular}
    \textsc{StarExp($\tau$)} is solvable in $(k\tau)!(k\tau)^{O(1)}$ time when the temporal modular-width of the graph is at most $k$.
\end{theorem}
\begin{proof}
    Given that the temporal modular-width of the graph is at most $k$, we check whether the number of leaves is more than $(k-1)\lfloor \tau / 2 \rfloor$. If the number of leaves is at least $(k-1)\lfloor \tau / 2 \rfloor + 1$ then, by the pigeon-hole principle and Lemma \ref{lem:star-group-leaves}, there must be at least $\lfloor \tau / 2 \rfloor + 1$ leaves whose edges to the central vertex are active at exactly the same times.  In this case, by Lemma \ref{lem:star-no}, we conclude that we have a no-instance of \textsc{StarExp($\tau$)}.
    
    Else, we have at most $(k-1)\lfloor \tau / 2 \rfloor < k \tau$ leaves. Note that, given an ordering of leaves to visit, we can check whether such a walk is valid in time polynomial in $k$ and $\tau$: we need only check for each leaf in order whether there are two appearances of its incident edge following the time-edges used to visit the previous leaf (and if so, greedily use the first two such appearances). Since there are fewer than $(k\tau)!$ possible orderings of the leaves, we can check each possibility in turn in time $(k\tau)!(k\tau)^{O(1)}$. 
\end{proof}

\section{Tractability with respect to Temporal Neighbourhood Diversity}\label{sec:TND}

We now turn our attention to our final parameter, temporal neighbourhood diversity, which is the most restrictive and hence allows for the most problems to be solved efficiently.  In Section \ref{sec:graph-burning} we demonstrate that \textsc{Temporal Graph Burning} is solvable in polynomial time when the temporal neighbourhood diversity is bounded by a constant (in fact we give an fpt-algorithm with respect to this parameterisation), even though the problem remains NP-hard when restricted to temporal graphs with constant temporal modular-width.  To illustrate further techniques that may be used to design efficient algorithms on graphs of bounded temporal neighbourhood diversity, in Section \ref{sec:min-edge-del} we also give an fpt-algorithm for the problem \textsc{SingMinReachDelete} with a single source vertex.

We begin with the formal definition of temporal neighbourhood diversity.  Once again, the definition is modelled on that for static graphs, which was first introduced by Lampis \cite{lampis_algorithmic_2012} and adapted by Ganian \cite{ganian_using_2012} to describe uncoloured graphs. In a static graph, we define the \emph{neighbourhood} $N(v)$ of a vertex $v$ as the set of vertices which share an edge with $v$. 
\begin{definition}[Type, Definition 2.2 \cite{ganian_using_2012}]
    Two vertices $v$, $v'$ have the same \emph{type} if and only if $N(v)\setminus\{v'\}=N(v')\setminus\{v\}$.
\end{definition}
\begin{definition}[Neighbourhood Diversity, Definition 2 \cite{lampis_algorithmic_2012}]
    A graph $G=(V,E)$ has \emph{neighbourhood diversity} at most $k$ if and only if there exists a partition of $V(G)$ into at most $k$ sets where, in each set, all vertices have the same type. We refer to this partition as a \emph{neighbourhood partition}.
\end{definition}

We note that the neighbourhood diversity of a graph can be computed in linear time \cite{lampis_algorithmic_2012}.

We now define the analogous temporal parameter, where we require that the edges between sets are all active at the same times.
\begin{definition}[Temporal Neighbourhood]
    The \emph{temporal neighbourhood} of a vertex $v$ in a temporal graph $(G, \lambda)$ is the set $TN(v)$ of vertex time pairs $(w, t)$ where $(w, t) \in TN(v)$ if and only if $vw \in E(G)$ and $t \in \lambda(vw)$.
\end{definition}

\begin{definition}[Temporal Type]
    Two vertices $u$, $v$ have the same \emph{temporal type} if and only if $\{(w,t) \in TN(u): w \neq v\} = \{(w,t) \in TN(v): w \neq u\}$. 
\end{definition}

\begin{definition}[Temporal Neighbourhood Diversity]
    A graph $\mathcal{G}$ has \emph{temporal neighbourhood diversity} at most $k$ if and only if there exists a partition of $V(\mathcal{G})$ into at most $k$ sets where, in each set, all vertices have the same temporal type. We refer to this partition as a \emph{temporal neighbourhood partition}. 
\end{definition}

It is immediate from this definition that, when all edges are assigned the same times, the temporal neighbourhood diversity of the graph is the same as the neighbourhood diversity of the underlying graph.  We now argue that, as in the static case, the subgraph induced by any class must form a clique or independent set; moreover, in the temporal setting, this must be true at every timestep.

\begin{lemma}\label{lem:class-ind-clique}
    At any snapshot $\mathcal{G}_t$ of $\mathcal{G}$, the subgraph induced by the vertices in a class $X$ of a temporal neighbourhood partition of $\mathcal{G}$ either forms an independent set or a clique.
\end{lemma}
\begin{proof}
    Let $u$ and $v$ be vertices in $X$, a class of a temporal neighbourhood partition of $\mathcal{G}$. Suppose the time-edge $(uv,t)$ exists in $\mathcal{G}$, then for any other vertex $w$ in $X$, the time-edges $(wu,t)$ and $(wv,t)$ must exist. Thus, the vertices in $X$ form a clique at time $t$.
    
    Similarly, suppose the time-edge $(uv,t)$ does not exist in $\mathcal{G}$. Then, for any other $w\in X$, the time-edges $(wu,t)$ and $(wv,t)$ cannot exist. Otherwise, $w$ has a different temporal neighbourhood to $v$ and $u$ respectively. Therefore, the vertices in $X$ must form an independent set at time $t$.
\end{proof}

As stated, temporal neighbourhood diversity is the most restrictive parameter in our hierarchy. Observe that for any temporal graph $\mathcal{G}$ the temporal modular-width is upper bounded by the temporal neighbourhood diversity: each class in the temporal neighbourhood partition forms a module, and it follows from Lemma~\ref{lem:class-ind-clique} that each of these modules can be constructed by operations~\ref{step1temp},~\ref{step2temp}, and~\ref{step3temp} of the temporal modular-width construction.

Finally, we argue that we can compute the temporal neighbourhood diversity efficiently. 

\begin{prop}\label{prop:poly-tnd}
    The temporal neighbourhood diversity of a temporal graph $(G,\lambda)$ can be calculated in $O(\Lambda n^3)$ time.
\end{prop}
\begin{proof}
    Observe that we can check in time $O(n\Lambda)$ whether two vertices are in the same class of a maximal temporal neighbourhood partition. There are at most $O(n^2)$ pairs of vertices to compare. Therefore, dividing vertices of a temporal graph into these equivalence classes can be done in $O(\Lambda n^3)$ time.
\end{proof}

\subsection{Temporal Graph Burning}\label{sec:graph-burning}

In this section we define \textsc{Temporal Graph Burning}, a temporal analogue of the static \textsc{Graph Burning} problem first proposed by Bonato et al.~\cite{DBLP:conf/waw/BonatoJR14}. Informally, a fire is placed at a new vertex in each time-step and the fire spreads along incident active edges. The problem asks if, after $h$ placements of fire, all vertices are burning. Static \textsc{Graph Burning} is NP-hard on general graphs \cite{DBLP:journals/dam/BessyBJRR17} and was recently shown to be in FPT parameterised by static modular width \cite{DBLP:journals/algorithmica/KobayashiO22}. In contrast, we prove that \textsc{Temporal Graph Burning} remains NP-hard on graphs with constant temporal modular-width.  This difference arises from the fact that, in the static setting, the length of a longest induced path in the graph (which is upper bounded by the modular-width) gives an upper bound on the time taken to burn the graph.  In the temporal setting, on the other hand, the times assigned to edges mean that even graphs with small diameter may take many steps to burn. In contrast, we show that \textsc{Temporal Graph Burning} can be solved in time $O(n^5\Lambda k!4^k)$ on temporal graphs with $n$ vertices, lifetime $\Lambda$ and temporal neighbourhood diversity $k$.

The \textsc{Temporal Graph Burning} problem asks how quickly a fire can be spread over the vertices of a temporal graph in the following discrete time process, where a fire is placed at a vertex of a graph in each timestep.
\begin{enumerate}
    \item At time $t = 0$ a fire is placed at a chosen vertex. All other vertices are unburnt.
    \item At all times $t \geq 1$, the fire spreads, burning all vertices $u$ adjacent to an already burning vertex $v$ where the edge between $u$ and $v$ is active at time $t$. Then, another fire is placed at a chosen vertex.
    \item This process ends once all vertices are burning.
\end{enumerate}

We refer to a sequence of vertices at which fires are placed as a strategy. 

\begin{definition}[Burning Strategy]
    A \emph{burning strategy} for a temporal graph $(G, \lambda)$ is a sequence of vertices $S = s_0, s_1, ..., s_\ell$ such that $s_i \in V(G)$ for all $i \leq \ell$, and in each timestep $i$ a fire is placed at $s_i$.
\end{definition}

We say that a strategy $S = s_0, s_1, ..., s_h$ has length $h$. For convenience, we allow for strategies that place fires at already burning vertices, although it is worth noting that such moves may be omitted. If every vertex in the graph is burning after a strategy is played, we say that strategy is successful.

\begin{definition}[Successful Burning Strategy]
\label{def:burnstrat}
Let $S = s_0, s_1, s_2, ..., s_\ell$ be a burning strategy for a temporal graph $(G, \lambda)$. We say $S$ is \emph{successful} if every vertex in $G$ is burning in timestep $\ell$ when the moves from $S$ are played.
\end{definition}

The decision problem asks how many timesteps it takes to burn a given temporal graph.

\decisionproblem{Temporal Graph Burning}{A temporal graph $(G, \lambda)$ and an integer $h$.}{Does there exist a  successful burning strategy for $(G, \lambda)$ of length less than or equal to $h$?}

This problem is in NP, with a strategy providing a certificate. Given a strategy it can be checked in polynomial time if it is successful and of length less than or equal to $h$ by simulating temporal graph burning on the input graph.

We show that \textsc{Temporal Graph Burning} is NP-hard even on graphs of bounded temporal modular-width. This is achieved by reducing from \textsc{(3, 2B)-SAT}, an NP-hard variant of the Boolean satisfiability problem in which each variable appears exactly twice both positively and negatively \cite{DBLP:journals/eccc/ECCC-TR03-049}, defined formally as follows. 

\decisionproblem{(3, 2B)-SAT}{A pair $(B, C)$ where $B$ is a set of Boolean variables, and $C=C_1 \land ... \land C_m$ is a set of clauses over $B$ in CNF, each containing 3 literals $C_j^1 \lor C_j^2 \lor C_j^3$, such that each variable appears exactly twice negatively and exactly twice positively.}{Is there a truth assignment to the variables such that all of the clauses in $C$ are satisfied?}

Our reduction produces a graph where each edge is active on exactly one timestep, and furthermore every connected component has a bounded temporal neighbourhood diversity, and hence the graph has bounded temporal modular-width by our earlier observation.

\begin{theorem}
\textsc{Temporal Graph Burning} is NP-hard even when restricted to graphs with constant temporal modular-width.
\end{theorem}
\begin{proof}
    We reduce from \textsc{(3, 2B)-SAT}. As such we begin by describing how to construct an instance $((G, \lambda), h)$ of \textsc{Temporal Graph Burning}, given an instance $(B, C)$ of \textsc{(3, 2B)-SAT} with $C=C_1 \land \dots \land C_m$, and $|B|=n$. We then go on to show that $(B, C)$ is a yes-instance of \textsc{(3, 2B)-SAT} if and only if $((G, \lambda), h)$ is a yes-instance of \textsc{Temporal Graph Burning}.

    To construct $((G, \lambda), h)$, begin by setting $h=2n+3m$. Now let the vertex set of $G$ be given by the union of the following sets:
    \begin{itemize}
        \item the set of \textit{literal vertices}: $\{x_i, \neg x_i: i \in [n]\}$,
        \item the set of \textit{clause vertices}: $\{c_j^1,c_j^2,c_j^3: j \in [m]\}$,
        \item the set of \textit{appearance vertices}: $\{u_{i,j},w_{i,j}: \text{either } x_i \text{ or } \neg x_i \text{ appears in } C_j\}$,
        \item the set of \textit{leaf vertices}: $\{ y_{i,d}, \lnot y_{i,d}: i \in [n], d \in [h+1] \} \cup \{ z_{j,d}^1, z_{j,d}^2, z_{j,d}^3: j \in [m], d \in [h+1] \}$, and
        \item a single isolated vertex $a$.
    \end{itemize}

    We represent the set of time-edges as a set of pairs of edges and a single timestep, such that if $(\{u,v\},t)$ is a time-edge then $\{u,v\}\in E(G)$ and $\lambda(\{u,v\})=\{t\}$. For clarity, each edge is active in exactly one timestep. This set is then given by the union of the following sets:

    \begin{itemize}
        \item $\{ (\{x_i,y_{i,d}\}, 2i), (\{\lnot x_i,\lnot y_{i,d}\}, 2i): i \in [n], d \in [h+1] \}$,
        \item $\{ (\{c_j^1,z_{j,d}^1\}, 2n+3j), (\{c_j^2,z_{j,d}^2\}, 2n+3j), (\{c_j^3,z_{j,d}^3\}, 2n+3j): j \in [m], d \in [h+1] \}$,
        \item $\{ (\{x_i,u_{i,j}\}, 2i-1), (\{u_{i,j},w_{i,j}\}, h), (\{w_{i,j},c_j^\ell\}, 2n+3j-1): x_i\text{ is the }\ell^\text{th}\text{ literal in }C_j \}$,
        \item $\{ (\{\lnot x_i,u_{i,j}\}, 2i-1), (\{u_{i,j},w_{i,j}\}, h), (\{w_{i,j},c_j^\ell\}, 2n+3j-1): \lnot x_i\text{ is the }\ell^\text{th}\text{ literal in }C_j \}$.
    \end{itemize}

    Note that this graph consists of $2n+1$ connected components, the vertex $a$, and each remaining component corresponding to a literal $x_i$ or $\lnot x_i$. We denote by $H_{i}$, and $\lnot H_i$ the connected components containing $x_i$ and $\lnot x_i$ respectively. One of these connected components can be seen in Figure~\ref{fig:ap:burnreductionconcom}.

    \begin{figure}[ht]
    \centering
    \includegraphics[width =\textwidth]{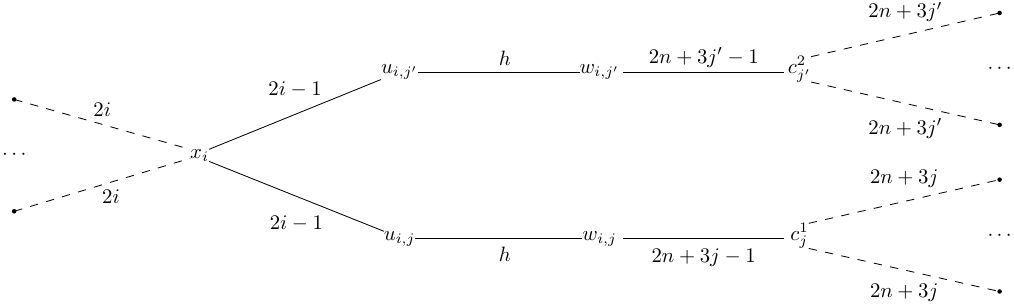}
    \caption{The connected component corresponding to the literal $x_i$, appearing in clauses $C_j$ and $C_{j'}$. Dashed edges indicate vertices and edges which are not pictured.}
    \label{fig:ap:burnreductionconcom}
\end{figure}

    In order to show that it is possible to burn this graph in $h$ timesteps if and only if there is a satisfying truth assignment for $(B, C)$, we first prove the following claims.

    \begin{claim}
        \label{clm:literalsinturn1}
        In order to burn $(G, \lambda)$ in $h$ or fewer timesteps, for each pair of literal vertices $x_i$ and $\lnot x_i$, a fire must be placed at or adjacent to one in timestep $2i-2$, and at the other in timestep $2i-1$.
    \end{claim}
    \begin{proof}
        Begin by observing that in order for $(G, \lambda)$ to burn in $h$ or fewer timesteps, every literal vertex $x_i$ or $\lnot x_i$ must be burning by the end of timestep $2i-1$, as every literal vertex has $h+1$ adjacent leaves with edges active at timestep $2i$. Observe that the fire can only spread to a literal vertex $x_i$ or $\lnot x_i$ by the end of timestep $2i-1$ if it originates at one of the two adjacent non-leaf vertices $u_{i,j}$ and $u_{i,j'}$. As a result, for each literal vertex $x_i$ or $\lnot x_i$ a fire must either be placed at the literal vertex by the end of timestep $2i-1$, or at one of the vertices $u_{i,j}$ and $u_{i,j'}$ by the end of timestep $2i-2$. We now continue by induction on the variable index $i$. For the base case it follows immediately that a fire must be placed at $x_1$ in timestep $0$ or $1$, or the adjacent vertex $u_{1,j}$ in timestep $0$. The same is true of the vertex $\lnot x_1$. Therefore in timestep $0$ a fire must be placed at or adjacent to one of these vertices, and in timestep $1$ a fire must be placed at the other vertex. For the inductive step assume that the claim is true for all literal vertices $x_a$ and $\lnot x_a$ with $a < i$. Then, in timestep $2i-2$, there must be no fires already placed either at or adjacent to $x_i$ or $\lnot x_i$. This follows from our inductive hypothesis that all earlier timesteps have been used to place fires on or next to literal vertices with lower indices than $i$. Then, a fire must be placed at $x_i$ in timestep $2i-2$ or $2i-1$, or at the adjacent vertex $u_{i,j}$ in timestep $2i-2$. Again, the same is true of vertex $\lnot x_i$, and therefore a fire must be placed at one of these vertices in timestep $2i-1$, and at or adjacent to the other in timestep $2i-2$.
    \end{proof}
    \begin{claim}
        \label{clm:clausesinturn1}
    	In order to burn $(G, \lambda)$ in $h$ or fewer timesteps, for each triple of clause vertices $c_j^1, c_j^2, c_j^3$ there must exist a permutation $\pi$ of $\{1, 2, 3\}$, such that a fire is placed at or adjacent to $c_j^{\pi(1)}$ in timestep $2n+3j-3$, at or adjacent to $c_1^{\pi(2)}$ in timestep $2n+3j-2$, and at $c_1^{\pi(3)}$ in timestep $2n+3j-1$.
    \end{claim}
    \begin{proof}
    	Begin by observing that in order for $(G, \lambda)$ to burn in $h$ or fewer timesteps, every clause vertex $c_j^\ell$ must be burning by the end of timestep $2n+3j-1$, as every clause vertex has $h+1$ adjacent leaves with edges active at timestep $2n+3j$. Observe that the fire can only spread to a clause vertex $c_j^\ell$ by the end of timestep $2n+3j-1$ if it originates at the adjacent non-leaf vertex $w_{i,j}$. As a result, for every clause vertex $c_j^\ell$ a fire must either be placed at the clause vertex by the end of timestep $2n+3j-1$, or at $w_{i,j}$ before the end of timestep $2n+3j-2$. We now continue by induction on the clause index $j$. For the base case we see from Claim~\ref{clm:literalsinturn1} that no fire can be placed at either $c_1^1$ or $w_{i,1}$ prior to timestep $2n$, and it therefore follows that a fire must be placed at $c_j^1$ in one of the timesteps $2n, 2n+1$, or $2n+2$, or the adjacent vertex $w_{i,1}$ in timesteps $2n$ or $2n+1$. The same is true of the two other clause vertices $c_1^2$ and $c_1^3$. Therefore there must exist a permutation $\pi$ of $\{1, 2, 3\}$, such that a fire is placed at or adjacent to $c_1^{\pi(1)}$ in timestep $2n$, at or adjacent to $c_1^{\pi(2)}$ in timestep $2n+1$, and at $c_1^{\pi(3)}$ in timestep $2n+2$. For the inductive step assume that the claim is true for all clause vertices $c_a^1, c_a^2, c_a^3$ with $a < j$. Then in timestep $2n+3j-3$, there must be no fires already placed either at or adjacent to $c_j^1$. Then, a fire must be placed at $c_j^1$ in one of the timesteps $2n+3j-1, 2n+3j-2, 2n+3j-3$, or at the adjacent vertex $w_{i,j}$ in one of the timesteps $2n+3j-2, 2n+3j-3$. The same is true of the vertices $c_j^2$ and $c_j^3$, and therefore there must exist a permutation $\pi$ of $\{1, 2, 3\}$, such that a fire is placed at or adjacent to $c_j^{\pi(1)}$ in timestep $2n+3j-3$, at or adjacent to $c_1^{\pi(2)}$ in timestep $2n+3j-2$, and at $c_1^{\pi(3)}$ in timestep $2n+3j-1$.
    \end{proof}

    \begin{claim}
        \label{clm:satclause1}
        Suppose that there is a successful strategy for $(G,\lambda)$ of length at most $h$, and consider the connected component $H_i$ or $\lnot H_i$ in which a fire is placed in timestep $2i-1$. Let $c_j^{\ell}$ be a clause vertex in this connected component. Then a fire must be placed at or adjacent to $c_j^{\ell}$ prior to timestep $2n+3j-1$.
    \end{claim}
    \begin{proof}
        If a fire is placed in $H_i$ or $\lnot H_i$ in timestep $2i-1$, then it must be placed at the corresponding literal vertex $x_i$ or $\lnot x_i$ by Claim~\ref{clm:literalsinturn1}. Note that, if the fire were to be placed on a non-leaf neighbour of $x_i$ (respectively $\lnot x_i$) at this time, the fire would not spread to $x_i$ (resp. $\lnot x_i$) because the fire would be placed after the edge between them is active. Therefore, we can assume that the fire is placed at the corresponding literal vertex (rather than its neighbour) at time $2i-1$.
        The fire cannot spread to the adjacent non-leaf vertex $u_{i,j}$ from the literal vertex, as the edge between $u_{i,j}$ and the literal vertex is only active at time $2i-1$. Thus, the fire must spread to $u_{i,j}$ from $w_{i,j}$. And, by Claim~\ref{clm:clausesinturn1}, a fire is either placed at $c_j^\ell$ one of the timesteps $2n+3j-3, 2n+3j-2, 2n+3j-1$ or at $w_{i,j}$ in one of the timesteps $2n+3j-3, 2n+3j-2$. This fire cannot be placed at $c_j^\ell$ in timestep $2n+3j-1$, as otherwise it would not reach $w_{i,j}$. In every other case, the fire is placed prior to timestep $2n+3j-1$ as required.
    \end{proof}

    \begin{claim}\label{clm:iso}
        Suppose there is a successful strategy for $(G,\lambda)$ of length at most $h$. Then a fire must be placed at $a$ in timestep $h$.
    \end{claim}
    \begin{proof}
        Since $a$ is an isolated vertex, the only way for it to burning is for a fire to be placed at $a$. In Claims~\ref{clm:literalsinturn1},~\ref{clm:clausesinturn1} and~\ref{clm:satclause1}, we prescribe the placement of fires in the first $2n+3m-1$ timesteps. This leaves only the placement of the fire in timestep $h=2n+3m$, which must be places at $a$ for a successful strategy.
    \end{proof}
    
    We are now ready to prove the correctness of our reduction, and begin by showing that if $((G, \lambda), h)$ is a yes-instance, then so is $(B,C)$.
    
    Assume that $((G, \lambda), h)$ is a yes-instance, and consider a strategy that burns the graph in $h$ or fewer timesteps. We then assign a value to each variable $x_i$ according to when the strategy places fires in the connected components $H_i$ or $\lnot H_i$. If in timestep $2i-2$ a fire is placed in $H_i$, then $x_i$ is assigned true, otherwise, if a fire is placed in $\lnot H_i$ then $x_i$ is assigned false. Note that by Claim~\ref{clm:literalsinturn1} this truth assignment is well defined. Now consider any clause $C_j \in C$, and see that by Claim~\ref{clm:clausesinturn1} one of the corresponding vertices $c_j^1, c_j^2$, or $c_j^3$ must have a fire placed at it in timestep $2n+3j-1$. Then, by Claim~\ref{clm:satclause1}, this vertex must be in the connected component in which a fire was placed in timestep $2i-2$. This literal is therefore assigned true, and thus the clause is satisfied, as required.
    
    Now assume that $(B, C)$ is a yes-instance, and consider a satisfying assignment of the variables in $B$. We now describe a burning strategy for $(G, \lambda)$, and show that it burns the graph in $h$ or fewer timesteps. We place fires such a fire is placed at the vertex corresponding to the truthful literal in $\{x_i,\lnot x_i\}$ in timestep $2i-2$, and a fire is placed at the vertex corresponding to the literal assigned false in $\{x_i,\lnot x_i\}$ in timestep $2i-1$. As we have a satisfying assignment, each clause $C_j = C_j^1 \lor C_j^2 \lor C_j^3$ must contain a literal $C_j^\ell$ that evaluates to true, and we place a fire at the corresponding clause vertex $c_j^\ell$ in timestep $2n+3j-1$. Fires are placed at the other two clause vertices $\{ c_j^1, c_j^2, c_j^3 \} \setminus \{ c_j^\ell \}$ in timesteps $2n+3j-3$ and $2n+3j-2$ in either order. Finally, a fire is placed at the vertex $a$ in timestep $h$. In this strategy, every literal vertex $x_i$ or $\lnot x_i$ is burnt by the end of timestep $2i-1$, and therefore all literal leaves $y_{i,d}$ burn by the end of timestep $h$. Every clause vertex $c_j^\ell$ is burnt by the end of timestep $2n+3j-1$, and therefore all clause leaves $z_{j,d}^\ell$ burn by the end of timestep $h$. Finally, if a vertex pair $u_{i,j}$ and $w_{i,j}$ belongs to a connected component $H_i$ or $\lnot H_i$ in which a fire is placed in timestep $2i-2$, the fire spreads from the literal vertex to $u_{i,j}$ in timestep $2i-1$, and then from $u_{i,j}$ to $w_{i,j}$ in timestep $h$. Otherwise, the vertex pair must be on a path with a clause vertex at which a fire is placed by timestep $2n+3j-2$, then the fire spreads from the clause vertex to $w_{i,j}$ in timestep $2n+3j-1$, and from $w_{i,j}$ to $u_{i,j}$ in timestep $h$. Thus in either case both of these vertices will burn by the end of timestep $h$ as required.
\end{proof}

We now show that \textsc{Temporal Graph Burning} is solvable efficiently when the temporal neighbourhood diversity of the input graph is bounded. Throughout we assume that the lifetime $\Lambda$ of the input temporal graph is at most the number of vertices $n$, as it is possible to burn any temporal graph in $n$ timesteps by placing a fire at every vertex in turn.

We begin by defining notation for the burning set of vertices in a given timestep when a strategy is played.
\begin{definition}[Burning Set]
    Given a strategy $S$ the \emph{burning set} $B_{t}(S)$ at timestep $t$ is the set of vertices burning immediately after a fire is placed in timestep $t$ when $S$ is played.
\end{definition}

We now give a number of results about how successful strategies can be modified to have certain desirable properties; these results allow us to bound the number of possible strategies we must check to determine whether there is a successful strategy of the desired length.

We now prove a lemma that shows that if, for two strategies $S$ and $R$ and some timestep $t_1$, we have $B_{t_1}(S) \subseteq B_{t_1}(R)$, then we are able to use an initial segment from strategy $R$ and find remaining moves from times $t_1+1$ onwards to obtain a strategy that will burn the graph at least as fast as $S$.

Thus throughout, in order to show the existence of a strategy that burns the graph at least as fast as another, we need argue only about the existence of such a timestep $t$ and the first $t$ moves made by the strategies.

\begin{lemma}
\label{lem:settimelin1}
Let $S$ be a successful strategy for $(G, \lambda)$. Suppose that there is some timestep $t_1 < |S|$ and strategy $R$ with $|R|=|S|$ such that $B_{t_1}(S) \subseteq B_{t_1}(R)$ and in every timestep after $t_1$, $R$ places a fire at the same vertex as $S$. Then $R$ is also successful.
\end{lemma}
\begin{proof}
We prove by induction on $t_2-t_1$ that for any timestep $t_2 \geq t_1$ we have that $B_{t_2}(S) \subseteq B_{t_2}(R)$.

Our base case is given, as $R$ is defined such that $B_{t_1}(S) \subseteq B_{t_1}(R)$. We now assume that $B_t(S) \subseteq B_t(R)$, for some $t_1 \leq t < |S|$, and show that $B_{t+1}(S) \subseteq B_{t+1}(R)$.

Now let $v \in B_{t+1}(S)$. If $v \in B_t(S)$ then $v \in B_t(R)$ and therefore $v \in B_{t+1}(R)$. Otherwise if $v \notin B_t(S)$ then either the fire spreads to $v$ in timestep $t+1$ when $S$ is played, or $S$ places a fire at $v$ in timestep $t+1$. In the former case $v$ is temporally adjacent to a vertex in $u$ in $B_t(S)$ in timestep $t+1$. It must be the case that $u$ is also in $B_t(R)$, and therefore the fire will spread to $v$ at $t+1$ when $R$ is played if $v$ has not already burnt. In the latter case either $v$ is burning before $R$ places a fire in timestep $t+1$ when $R$ is played, or $r_{t+1} = s_{t+1} = v$.

Therefore $B_{t+1}(S) \subseteq B_{t+1}(R)$. We then have that $V(G) = B_{|S|}(S) \subseteq B_{|S|}(R)$, and therefore $R$ is successful.
\end{proof}

We now show that we can delay placing a fire at a vertex belonging to a class in which the fire is already burning, without causing a large effect on the set of vertices that will burn at each timestep.

\begin{lemma}
    \label{lem:alreadyburn1}
    Let $(G, \lambda)$ be a temporal graph with temporal neighbourhood partition $(X_i)_{i\in I}$. Now let $S$ be a strategy that burns this graph, and let $u$ be a vertex at which $S$ places a fire in a timestep $t_1$, and $X_i$ be the class to which this vertex belongs. Fix a timestep $t_2 > t_1$, and let $S'$ be a strategy which plays as follows until timestep $t_2$:

    \[
    s'_t=\begin{cases}
        s_t&\text{if }t<t_1\\
        s_{t+1}&\text{if }t_1\leq t < t_2\\
        u&\text{if }t=t_2
    \end{cases}
    \]

    If there exists a vertex $w \in X_i$ which is burning before the end of timestep $t_1$ when $S'$ is played, we have that $B_{t_2}(S) \subseteq B_{t_2}(S')$.
\end{lemma}
\begin{proof}
    We will show that for any vertex $x \in B_{t_2}(S)$ we have that $x \in B_{t_2}(S')$.

    Consider the case where $x$ is a vertex at which $S$ places a fire. If $x \neq u$ then $S'$ will place a fire at $x$ in the same timestep or earlier than $S$, and thus $x \in B_{t_2}(S')$. Otherwise, if $x = u$ then $S'$ places a fire at it in timestep $t_2$, and again $x \in B_{t_2}(S')$.

    Now, consider the case where $S$ does not place a fire at $x$ before timestep $t_2$. It must then burn because the fire spreads to it. Note, as $x \in B_{t_2}(S)$, there exist timesteps $t_\alpha$ and $t_\beta$ with $t_\alpha < t_\beta$ such that there exists a temporal path $P$ with arrival time $t_\beta$ that traverses the vertices $y_1, ..., y_p$, $y_p = x$, and $y_1$ is a vertex at which $S$ places a fire prior to $t_\alpha$, and $t_\beta \leq t_2$.

    If $y_1 \neq u$ then $x = y_p$ will still burn before the end of timestep $t_\beta \leq t_2$ when $S'$ is played, as $S'$ will also place a fire at $y_1$ prior to timestep $t_\alpha$. Thus $x \in B_{t_2}(S')$.

    Otherwise if $y_1 = u$ then $u$ is temporally adjacent to $y_2$ in timestep $t_\alpha > t_1$. As $w, u \in X_i$, $w$ is also temporally adjacent to $y_2$ in this timestep, and there is also a temporal path $P' = w, y_2, ..., y_p$ starting at time $t_\alpha > t_1$, and identical to $P$ in all but the first vertex. Therefore in this case when $S'$ is played $x = y_p$ will burn before the end of timestep $t$, as $w$ burns by the end of timestep $t_1 < t_\alpha$ when $S'$ is played. Thus, again, $x \in B_{t_2}(S')$.
\end{proof}

We go on to show that any time we place a fire at a vertex, we may instead place a fire at another unburnt vertex in the same class, if such a vertex exists, and obtain a strategy that burns the graph in the same time.

\begin{lemma}
    \label{lem:anyfromclass1}
    Let $(G, \lambda)$ be a temporal graph with temporal neighbourhood partition $(X_i)_{i\in I}$.

    Let $S$ and $S'$ both be strategies with $|S'| = |S|$, such that in every timestep, $S$ and $S'$ both place fires in the same class, that is, for any $i \leq |S|$, we have that there exists a class $X_j$ where $\{ s_i, s'_i \} \subseteq  X_j$. Furthermore, suppose that $S'$ places a fire at an already burning vertex in a timestep $i$ if and only if $S$ also places a fire at an already burning vertex in timestep $i$.

    Then $S$ is successful if and only if $S'$ is.
\end{lemma}
\begin{proof}
    We show that in each timestep $t$, the number of burning vertices in each class from the temporal neighbourhood partition is the same when $S$ and $S'$ are played. We proceed by induction on the timestep.

    In the first timestep only $s_1$ is burning when $S$ is played, and only $s'_1$ is burning when $S'$ is played. By assumption these two vertices are both in the same class in the temporal neighbourhood partition.

    Then, assume that in some timestep $t$ the number of burning vertices in each class from the temporal neighbourhood partition is the same when $S$ and $S'$ are played. Now given any class $X_i$ from the temporal neighbourhood partition, let $b_i$ be the number of vertices burning in $X_i$ at the end of timestep $t+1$ when $S$ is played. The number of vertices burning in $X_i$ at the end of timestep $t+1$ when $S'$ is played is then the number of vertices that were burning in timestep $t$, plus the number of vertices to which the fire spreads, plus one if a fire was placed in $X_i$ by $S'$ in timestep $t+1$. All of the vertices in $X_i$ will be burning by the end of timestep $t+1$ if the  fire spreads to any vertex $v \in X_i$, as any burning vertex $u$ adjacent to $v$ in $t+1$ is also adjacent to all other vertices in $X_i$. Furthermore, such a vertex $u \in X_j$ exists if and only if there is a burning vertex in $X_j$ at the end of timestep $t$ when $S$ is played, as the number of vertices burning in $X_j$ at the end of timestep $t$ is the same when both $S$ and $S'$ are played. Then, all of the vertices of $X_i$ are burning in timestep $t+1$ when $S$ is played if and only if all of the vertices of $X_i$ are burning in timestep $t+1$ when $S'$ is played. Furthermore, $S'$ places a fire in $X_i$ if and only if $S$ does also. Therefore, in timestep $t+1$ the number of vertices burning when $S$ is played is the same as the number of vertices burning when $S'$ is played.

    Then, by induction, if there exists a timestep $t$ in which all the vertices are burning when $S$ is played, the same number of vertices are burning when $S'$ is played, so $S$ is successful if and only if $S'$ is.
\end{proof}

\begin{definition}[Placement Classes]
   Given a strategy $S$, the set of classes from the temporal neighbourhood partition in which $S$ places fires is referred to as the placement classes $C(S)$ of $S$.
\end{definition}

We now show that we can reorder any successful burning strategy $S$, so that initially one fire is placed in every \textit{placement class} for $S$. This reordering gives a new strategy which is still successful.

\begin{lemma}
\label{lem:ddfirst1}
Given a temporal graph $(G, \lambda)$, let $S$ be any successful strategy. There is then a successful strategy $S'$ with $|S'| = |S|$, and $C(S') = C(S)$, such that the first $|C(S)|$ burns are in distinct equivalence classes in the temporal neighbourhood partition.
\end{lemma}
\begin{proof}
    Assume that $(G, \lambda)$ is a counterexample. Then let $R$ be a successful strategy minimal in the timestep $t_c$, such that at the end of timestep $t_c$ there is a fire placed in every class in $C(R)$.  Now, let $u \in X_i$ be a vertex at which $R$ places a fire in timestep $t_1 \leq |C(R)|$, such that a fire has already been placed at a vertex $w \in X_i$ prior to $t_1$. 

    Consider the strategy $R'$ which makes moves as follows:
    \[r'_t=\begin{cases}
        r_t &\text{if }t<t_1\text{ or }t>t_c\\
        r_{t+1} &\text{if } t_1 \leq t < t_c\\
        u &\text{if }t=t_c.
    \end{cases}\]

    By Lemma~\ref{lem:alreadyburn1}, we have that $B_{t_c}(R) \subseteq B_{t_c}(R')$. Then, by Lemma~\ref{lem:settimelin1}, $R'$ burns $(G, \lambda)$ in the same or less time as $R$. This contradicts the assumption that $R$ was minimal in the number of moves played until a fire is placed in every class in $C(R)$, so no such counterexample $(G, \lambda)$ can exist.
\end{proof}

Finally we show that, given a strategy $S$ that places fires only in distinct classes for the first $|C(S)|$ moves, we can arbitrarily reorder all subsequent moves made after timestep $|C(S)|$.
\begin{lemma}
    \label{lem:arbord1}
    Let $(G, \lambda)$ be a temporal graph, and $S$ a successful strategy such that the first $|C(S)|$ fires placed by $S$ are placed in distinct classes from the temporal neighbourhood partition. Let $f : [|C(S)|+1,|S|] \to [|C(S)|+1,|S|]$ be any bijection.

   Then the strategy $S'$ given by

    \[s'_t=\begin{cases}
        s_t&\text{if }t\leq|C(S)|\\
        s_{f(t)}&\text{otherwise}
    \end{cases}\]

    is successful.
\end{lemma}
\begin{proof}
    We begin by showing that $S'$ is in fact successful. We argue this by contradiction. Suppose there is a vertex $v$ which is not burning once all the moves in $S'$ are played. Then $v$ must be a vertex on which no fire is placed in either $S$ or $S'$. This implies that the fire spreads to $v$ from another vertex; denote by $w$ the vertex from which the fire spreads. Assume without loss of generality that $w$ is the first in its class to be burned. Then $w$ must be burned in the first $C(S)$ timesteps, and so the fire spreads to $v$ in both strategies. This contradicts the assumption that $v$ is never burned, showing that $S'$ is a successful strategy.





\end{proof}

We now present an algorithm for \textsc{Temporal Graph Burning}, and show that this algorithm is an fpt-algorithm with respect to temporal neighbourhood diversity.

\begin{algorithm}\caption{TND Graph Burning Algorithm}\label{alg:grap-burn-alg1}
    \begin{algorithmic}[1]
    \Require A temporal graph $\mathcal{G}$, and an integer $h$.
    \Ensure True if there exists a successful burning strategy of length at most $h$, and false otherwise.
    \State Compute the temporal neighbourhood partition $\Theta$ of $(G, \lambda)$.
    \ForAll{possible subsets $A \subseteq \Theta$}
        \ForAll{possible orderings of $A$}
            \ForAll{possible subsets $B \subseteq A$}
            \State Let $S$ be the strategy that first places a fire in order in every class from $A$, and then places fires at every unburnt vertex in $B$.
                \If{$S$ is successful and consists of $h$ or fewer moves} 
            \State \textbf{return} true.
            \EndIf
            \EndFor
        \EndFor
    \EndFor
    \State \textbf{return} false.
    \end{algorithmic}
\end{algorithm}

We now prove correctness of this algorithm, using the following lemmas.

\begin{lemma}
    Algorithm~\ref{alg:grap-burn-alg1} returns true for a temporal graph $(G, \lambda)$ and integer $h$ if and only if there exists a strategy $S$ that burns the graph in $h$ or fewer timesteps.
\end{lemma}
\begin{proof}
    Suppose there exists a strategy $S$ that burns $(G, \lambda)$ in $h$ or fewer timesteps. By Lemma~\ref{lem:ddfirst1} and Lemma~\ref{lem:anyfromclass1} we may assume without loss of generality that $S$ first places a fire at an arbitrary vertex from every class in $C(S)$. The remaining moves must then place fires at every other vertex of any class to which the fire will not spread before the graph is burnt. These classes must be some subset of the classes from $C(S)$, and from Lemma~\ref{lem:arbord1} we know that these fires can be placed in any order.

    The algorithm exhaustively checks every such strategy, and thus will return true if any strategy exists that burns $(G, \lambda)$ in $h$ or fewer moves, and false otherwise.
\end{proof}

This allows us to obtain fixed parameter tractability, as bounding the temporal neighbourhood diversity bounds the number of such strategies that we have to check.

\begin{theorem}
    \textsc{Temporal Graph Burning} is solvable in time $O(\Lambda n^3\phi\phi!2^\phi)$, where $n$ is the number of vertices in $\mathcal{G}$, $\Lambda$ the lifetime, and $\phi$ the temporal neighbourhood diversity. If the temporal neighbourhood partition is given, we obtain a runtime of $O(\Lambda n^2\phi\phi!2^\phi)$.
\end{theorem}
\begin{proof}
    The TND Graph Burning Algorithm solves \textsc{Temporal Graph Burning}. It begins by computing the temporal neighbourhood partition, which we know from Proposition~\ref{prop:poly-tnd} that we can do in time $O(\Lambda n^3)$. Furthermore, there are at most $\phi\phi!$ possible orderings $A$ of subsets of the temporal neighbourhood partition $\Theta$, and at most $2^\phi$ sets $B$. We then simulate temporal graph burning on the graph, which is possible in $O(n^2\Lambda)$ time, where $n$ is the number of vertices in the input graph. This gives us an overall running time of $O(\Lambda n^3 + \phi\phi!2^\phi n^2\Lambda)=O(\Lambda n^3\phi\phi!2^\phi)$. If the decomposition is given, we instead obtain a runtime of $O(\Lambda n^2\phi\phi!2^\phi)$, as we may drop the extra factor of $n$ needed to compute it.
\end{proof}

\subsection{Minimum Reachability Edge Deletion}\label{sec:min-edge-del}
Here we give another example of an algorithm which uses the temporal neighbourhood decomposition.  Loosely, the problem we are solving asks: given a specified source vertex, what is the minimum number of edge appearances that can be deleted to limit the number of vertices reachable from that source to at most some value?

We say a vertex $v$ is \emph{temporally reachable} from a vertex $u$ in $\mathcal{G}$ if there exists a temporal path from $u$ to $v$. We say a vertex $v$ is temporally reachable from a set $S$ if there is a vertex in $S$ from which $v$ is temporally reachable. The \emph{reachability set} $\text{reach}(v)$ of a vertex $v$ is the set of vertices temporally reachable from $v$. The \emph{temporal reachability} of a vertex is the cardinality of its reachability set.
We can now give the formal problem definition; this is a special case of the problem \textsc{MinReachDelete} studied by Molter et~al.~\cite{molter_temporal_2021} in which multiple sources are allowed.

\label{MinTRED}
\decisionproblem{Singleton Minimum Temporal Reachability Edge Deletion (SingMinReachDelete)}{A temporal graph $\mathcal{G}=(G,\lambda)$, a vertex $v_s \in V(G)$ and positive integer $r$.}{What is the cardinality of a smallest set of time-edges $E$ such that the vertex $v_s$ has temporal reachability at most $r$ after their deletion from $\mathcal{G}$?}

\textsc{SingMinReachDelete} was shown by Enright et al.~\cite{enright_deleting_2021} to be NP-hard (and W[1]-hard parameterised by the maximum number of vertices that are allowed to be reached following deletion) even when the lifetime of the input temporal graph is $2$ and every edge is active at exactly one timestep. In their problem, they ask if there exists a deletion such that no vertex in the resulting temporal graph reaches more than $r$ vertices.
While the result of \cite{enright_deleting_2021} is for a version of the problem when the source set $S$ is the entire vertex set, it is clear from the construction that hardness also holds with a single source vertex. 

We show that this problem is in FPT when parameterised by temporal neighbourhood diversity and the \emph{temporality} of the input graph $\tau(\mathcal{G})$, which was defined by Mertzios et al. \cite{mertzios_temporal_2019} to be the maximum number of times an edge appears.  When the temporal graph in question is clear from context, we just refer to $\tau$.  We note that it remains open whether the problem belongs to FPT parameterised by temporal neighbourhood diversity alone, or indeed parameterised by temporal modular width or temporal cliquewidth; the techniques we use here do not extend naturally to these less restrictive settings.

We now give a formal statement of our result.

\begin{theorem}\label{thm:iqp}
    \textsc{SingMinReachDelete} is solvable in time $g(k,\tau)\log^{O(1)}r + \Lambda n^3$, where $g$ is a computable function.  If a temporal neighbourhood decomposition is given, we can solve the problem in time $g(k,\tau)\log^{O(1)} r$.
\end{theorem}
This uses a result by Lokshtanov \cite{lokshtanov_parameterized_2017} which gives an FPT algorithm for the following problem.

\decisionproblemparam{Integer Quadratic Programming}{A $n\times n$ integer matrix $Q$, an $m\times n$ matrix $B$ and an $m$-dimensional vector $\textbf{b}$.}{$n+\alpha$ where $\alpha$ is the maximum absolute value of any entry in $B$ or $Q$.}{Find a vector $\textbf{x}\in \mathbb{Z}^n$ which minimises $\textbf{x}^TQ\textbf{x}$, subject to $B\textbf{x}\leq \textbf{b}$.}

Their result is as follows.
\begin{theorem}[Theorem 1, \cite{lokshtanov_parameterized_2017}]\label{thm:iqp-ref}
    There exists an algorithm that, given an instance of \textsc{Integer Quadratic Programming}, runs in time $g(n,\alpha)L^{O(1)}$ (for some computable function $g$), and outputs a vector $x\in\mathbb{Z}^n$. Here $L$ is the total number of bits required to encode the input integer quadratic program. If the input IQP has a feasible solution then $x$ is feasible, and if the input to the IQP is not unbounded, then $x$ is an optimal solution.
\end{theorem}

Let $\mathcal{G}$ be the temporal graph in the input of \textsc{SingMinReachDelete}. We denote by $\mathcal{G}_s$ the temporal neighbourhood partition graph of $\mathcal{G}$. This is a temporal graph where the classes of the temporal neighbourhood partition form the vertex set. We refer to nodes of $\mathcal{G}_s$ and vertices of $\mathcal{G}$ to differentiate between the two. An edge exists at time $t$ between nodes $A$ and $B$ in $\mathcal{G}_s$ if and only if there exist edges in $\mathcal{G}$ at time $t$ between (all) vertices in $A$ and (all) vertices in $B$. Note that, given two vertices of the same temporal type, their reachability sets must consist of the same vertices except for the vertices themselves; as a result, the reachability sets of vertices in a given class all have the same cardinality.  Moreover, all vertices of the same type are first reached from the source at the same time (where we say a vertex is ``first reached'' at time $t$ if the final time-edge in an earliest-arriving temporal path from the source is at time $t$).  

Our strategy for solving \textsc{SingMinReachDelete} is as follows.  We partition each class according to the time at which the vertices are first reached after the deletion of time-edges, and consider all possibilities for which of these subclasses are non-empty.  Given a function $\phi$ telling us which subclasses are non-empty, we argue that we can determine efficiently whether there is indeed a deletion such that precisely these subclasses are non-empty, and if so compute exactly the pairs of subclasses between which we must delete time-edges to achieve this.  For a fixed $\phi$, we then encode the problem as an instance of \textsc{Integer Quadratic Programming}, where the variables are the sizes of the subclasses and the objective function seeks to minimise the number of time-edges we must delete.

From now on we denote by $A_t$ the subclass of vertices in class $A$ that are first reached at time $t$ following the deletion of time-edges, with an additional subclass $A_{\infty}$ consisting of those vertices in $A$ not temporally reachable from the source.

We begin by making some assertions about these subclasses.

\begin{lemma}\label{cor:iqp-number-subclasses}
    For each class $A$ in the temporal neighbourhood partition which does not contain the source, the number of indices $t$ such that $A_t$ can be non-empty following a deletion is at most $\tau (k+1)$ where $k$ is the temporal neighbourhood diversity of $\mathcal{G}$.
\end{lemma}
\begin{proof}
    It suffices to show that there are at most $\tau \cdot k$ possibilities for the earliest arrival time at any given vertex, since the set of possible earliest arrival times will be the same for all vertices in the same class; we then have one additional subclass for the ``unreached'' subclass.

    Suppose that $v$ is first reached from the source at time $t$ in $\mathcal{G}\setminus E$, for some set of time-edges $E$. Then there is a temporal path from the source to $v$ whose last time-edge is at time $t$, so in particular we know that there is an edge incident with $v$ in G that is active at time $t$.  The number of distinct times at which there is an edge incident with any vertex is at most $\tau\cdot k$, since there can be at most $\tau$ distinct times at which edges in that class and edges to each other class are active, giving the result.
\end{proof}

In total, therefore, we need to consider at most $k(k+1)\tau+1$ subclasses, where the additional subclass is the subclass containing only the source indexed with time 0.  Let $\phi(A,t)$ be a function on this set of subclasses which maps $A,t$ to the label empty to indicate that $A_t$ should be empty, and to non-empty otherwise.  
We now make some observations about which deletions of time-edges can reduce the cardinality of the temporal reachability set of the source.
\begin{lemma}\label{lem:iqp-useful-del}
    Let $e$ be an edge with endpoints $u$ and $v$. If neither $u$ nor $v$ is reached by time $t_1$ from the source, deletion of any appearances of $e$ at or before $t_1$ cannot change the reachability set of the source. Further, if both $u$ and $v$ are reached from the source by time $t_2$, deletion of any appearances of $e$ at or after time $t_2$ will not change the reachability set of the source.
\end{lemma}
\begin{proof}
    We begin by supposing that neither $u$ nor $v$ are reached from the source by time $t_1$. Then, if deletion of the edge $e$ at time $t_1$ were to change the reachability of the source, there would be a temporal path from the source to an endpoint of $e$ arriving strictly before $t_1$; a contradiction of our assumption.

    Now consider a vertex $w$ reachable from the source by a path $p$ which uses $e$ at time $t_2$. Suppose without loss of generality that, in $p$, the edge is traversed from $u$ to $v$ at time $t_2$. 
    Since $v$ is reachable from the source by time $t_2$, there must exist an alternative path from the source to $w$ which does not use the time-edge $(e,t_2)$. This is found by replacing the subpath of $p$ from the source to $v$ with the path from the source to $v$ which arrives by $t_2$. Therefore, we can still reach $w$ from the source without $e$ at time $t_2$, and deletion of $(e,t_2)$ does not change the reachability set of the source.
\end{proof}
\begin{corollary}\label{cor:internal-del}
    An optimal deletion will never include the deletion of a time-edge with both endpoints in the same subclass.
\end{corollary}
\begin{lemma}\label{lem:iqp-full-del}
     Let $\mathcal{G}'$ be the temporal graph obtained from making an optimal deletion of time-edges in $\mathcal{G}$. Fix $v$ to be any vertex, and $A$ to be any class of the temporal neighbourhood partition of $\mathcal{G'}$ which is reached by the source before $v$ is reached.  Then, in any snapshot of $\mathcal{G}'$, $v$ is either adjacent to all vertices in $A$ or none of them.
\end{lemma}
\begin{proof}
    Suppose for contradiction that, after an optimal deletion, the number of edges between $v$ and $A$ at time $t$ is $|E(v,A, t)|$, where $0<|E(v,A)|<|A|$. That is, it is neither complete nor empty.

    Let $w$ and $w'$ be vertices in $A$ such that the time-edge $(vw',t)$ is deleted and $(vw,t)$ is not after an optimal deletion. Then, since all of the vertices in $A$ are reached at a time before $t$ by the source, reinstating the time-edge $(vw',t)$ does not increase the reachability of the source. Thus, the deletion is not optimal; a contradiction. Hence, in an optimal deletion, the edges at some time $t$ are either complete or empty between a vertex and every subclass reached before its own.
\end{proof}

\begin{lemma}\label{lem:iqp-full-del-symetric}
    Let $\mathcal{G}'$ be the temporal graph obtained from making an optimal deletion of time-edges in $\mathcal{G}$, and let $A_i$ and $B_j$ be two subclasses.  In each snapshot of $\mathcal{G}'$, the graph is either complete or empty between $A_i$ and $B_j$. Therefore, all vertices in the same subclass have the same temporal neighbourhood in $\mathcal{G}'$.
\end{lemma}
\begin{proof}
    Suppose for a contradiction that there is a time $t$ at which the graph between $A_i$ and $B_j$ is neither complete nor empty. Let $i$ and $j$ be the respective times at which $A_i$ and $B_j$ are first reached in $\mathcal{G}'$.
 Denote by $E$ the set of edges deleted between $A_i$ and $B_j$ at time $t$ under this deletion. Without loss of generality, assume $j\geq i$. If $t\leq i$ or $t\geq j$, then by Lemma \ref{lem:iqp-useful-del}, the edges between $A_i$ and $B_j$ at time $t$ are not deleted in an optimal deletion.
 
 Therefore, we can assume that $i< t< j$.
 If there is at least one edge remaining between $A_i$ and $B_j$ then at least one vertex in $B_j$ must first be reached from the source at time $t<j$ by a path traversing an edge from a vertex in $A_i$ to a vertex in $B_j$ at time $t$. This contradicts our assumption that the vertices in $B_j$ are first reached at time $j$.
\end{proof}
The previous three lemmas describe which deletions are useful for reducing the reachability of the source and achieving some assignment $\phi$. We now specify the time needed to determine whether a given assignment $\phi$ is feasible and, if so, find the unique minimal set of times at which we must delete edges between each pair of subclasses in order to realise $\phi$.
\begin{lemma}\label{lem:iqp-possible-phi}
    Given an assignment $\phi$ of subclasses to empty and non-empty, we can check in time $O(k^5\tau^4)$ whether it is possible to delete time-edges from $\mathcal{G}$ so that, for each subclass $A$ and index $t$, $A_t$ is empty if and only if $\phi(A,t)=\text{empty}$. If such a deletion is possible, the algorithm outputs the unique minimal set of times at which we must delete edges between each pair of subclasses in order to realise $\phi$.
\end{lemma}
\begin{proof}
    Given $\phi$, we create a temporal graph $\mathcal{H}$ consisting of a node for each subclass of $\mathcal{G}$. Two nodes of $\mathcal{H}$ are adjacent at time $t$ if and only if the corresponding classes of $\mathcal{G}$ are adjacent at time $t$. Note that if two nodes of $\mathcal{H}$ are subclasses of the same class, they are adjacent at time $t$ if and only if there exist edges at that time in that class. Each node of $\mathcal{H}$ is labelled with the time at which it is first reached from the source. In the case of an ``unreached'' subclass, this time label is $\infty$. In the algorithm where we check whether an assignment $\phi$ is valid, we will refer to the nodes as pairs $(a,t)$ where $a$ is the name of the node and $t$ is the time at which it is first reached. By Lemma \ref{cor:iqp-number-subclasses} and the discussion following it, there are at most $k(k+1)\tau+1$ nodes in $\mathcal{H}$.
    
    Recall that $\phi$ is a function which takes each subclass to either an empty or non-empty marker. We check whether is is possible to realise assignment $\phi$ by a variation of breadth-first search (BFS) on the graph $\mathcal{H}$. We begin by finding the subgraph induced by removing any of the subclasses which are empty under $\phi$ from $\mathcal{H}$. Call this graph $\mathcal{H}'$. If the resulting graph is disconnected and there is a node with time $t\neq\infty$ in a different connected component of the underlying graph of $\mathcal{H}$ to the source, we reject. 

    We keep track of two values for each node $a$ in $\mathcal{H}$; $t^a_{\text{target}}$ and $t_{\text{first}}^a$. For the node corresponding to the subclass containing the source, we initialise both $t_{\text{first}}^a$ and $t_{\text{target}}^a$ to $0$. We initialise $t_{\text{first}}^a$ with the value $\infty$ for all remaining nodes $a$ in $\mathcal{H}'$ and set $t_{\text{target}}^a$ to be the time at which vertices in node $a$ should be first reached by the source, as prescribed by $\phi$. The value $t_{\text{first}}^a$ will be the earliest time of arrival from the source once the algorithm finishes. Therefore we accept if, for all nodes $a$, $t_{\text{target}}^a=t_{\text{first}}^a$ when the algorithm has finished executing. We initialise an empty set $E$ of edges which we will mark as deleted. We begin by adding the node containing the source to the queue. After dequeuing a node $a$, for each of its neighbours $b$, we check if the edge $ab$ is labelled with a time $t$ after $t_{\text{first}}^a$ and before $t_{\text{first}}^b$. For all $t$ after $t_{\text{first}}^a$ such that, $(ab,t)$ is in $\mathcal{H}'$, $(ab,t)$ is not in $E$, and $t$ is before $t_{\text{target}}^b$, we add $(ab,t)$ to $E$. Then we are left with two cases, either there is no time $t^*$ such that $ab$ is active at time $t^*$ where $t^*$ is after $t_{\text{first}}^a$ and before $t_{\text{first}}^b$, or such a time exists. If the former is true, we simply consider the next node in the queue. Else, $t^*$ must be equal to or after $t_{\text{target}}^b$. Let $t^*$ be the earliest time after $t_{\text{first}}^a$ such that $ab$ is active. Then, we update $t_{\text{first}}^b$ to $t^*$ and add $b$ to the queue. The algorithm terminates when the queue is empty.

    We note that a node can be added to the queue at most as many times as it could be reached from the source. This is $\tau \cdot k$ times. Furthermore, the operations done when considering a dequeued node take linear time in the number of nodes and temporality. By Lemma \ref{cor:iqp-number-subclasses}, there are at most $k(k+1)\tau$ nodes in $\mathcal{H}$. Therefore, the algorithm runs in $O(k^5\tau^4)$ time.

    Once the algorithm terminates, $t^a_{\text{first}}$ is the earliest time of arrival of a path from the source to the node $a$ given the deletions made. We make a deletion if and only if otherwise there is a temporal path from the source that arrives at a node $a$ before $t^a_{\text{target}}$. Therefore, $t^a_{\text{target}}=t^a_{\text{first}}$ for all nodes $a$ (and the algorithm accepts) if and only if there is a set of deletions $E$ such that the earliest time of arrival at a node in $\mathcal{H}$ is as prescribed by $\phi$. Observe that $(ab,t)$ is added to $E$ only if all edges at time $t$ between the subclasses corresponding to $a$ and $b$ must be deleted in order to realise $\phi$. Therefore $E$ describes the minimal set of times at which we must delete edges between each pair of subclasses.
\end{proof}

We can now transform our edge deletion problem given some $\phi$ into an instance of \textsc{IQP} and use the algorithm of Theorem \ref{thm:iqp-ref} to solve it efficiently.
\begin{lemma}\label{lem:iqp-cardinality-min}
    Given a fixed, feasible $\phi$ we can find in time $g(k,\tau)\log^{O(1)} r$ the cardinalities of the subclasses which minimise the number of deleted edges such that the source reaches $r$ vertices.
\end{lemma}
\begin{proof}
    Given a $\phi$, we can express the division of vertices among non-empty subclasses such that at most $r$ vertices are temporally reachable from the source and the minimum number of deletions are made as an instance of \textsc{Integer Quadratic Programming}. We arbitrarily order the subclasses. Then, the vector of variables is $\textbf{x}=(x_1,\ldots,x_{\ell})^T$, where the $i$th variable is the number of vertices in the $i$th subclass. The number of variables $\ell$ is at most $k(k+1)\tau$. Note that the subclass containing the source is not counted here, as we always know that this subclass has exactly one vertex in it. For a given assignment $\phi$, we can find the minimum deletion between the subclasses such that the assignment holds by Lemma \ref{lem:iqp-possible-phi}. Then, the matrix $Q_{\phi}$ is a $\ell\times \ell$ triangular matrix where there is a $y$ in position $q_{i,j}$ if we have determined that there are exactly $y$ timesteps at which we must delete all time-edges between subclass $i$ and subclass $j$ to match the assignments of $\phi$. Therefore, $x^TQx$ gives the number of edges we need to delete. We write $x^{A,t}$ for the entry of $\textbf{x}$ corresponding to the subclass of $A$ that is first reached from the source at time $t$.

    We employ the following linear constraints
    \begin{align}
        \nonumber \forall A,t,\,x^{A,t}\geq 0\\
        \sum_{t\neq \infty}x^{A,t}\leq r-1 \\
        \forall A,\, \sum_t x^{A,t}=|A|.
    \end{align}
The constraints ensure that every $x^{A,t}$ is non-negative and
\begin{enumerate}
    \item $\text{reach}(v_s)\leq r$, and
    \item the sum of cardinalities of the subclasses of $A$ is the cardinality of $A$.
\end{enumerate}
We can express these conditions in the form $B\textbf{x}\leq\textbf{b}$, where $B$ is a $(\ell+1+2k)\times \ell$ matrix and $b$ is a $(\ell +1+2k)$-dimensional vector.  Note that the largest absolute value in $Q_{\phi}$ is at most $\tau$ and the largest absolute value in $B$ is $1$. Recall that the number of variables $\ell$ is at most $k(k+1)\tau$. By Theorem \ref{thm:iqp-ref}, we can solve this instance of \textsc{IQP} in time $f(k,\tau)L^{O(1)}$ where $L$ is the number of bits required to encode the input integer quadratic program.
Combining the parts dependent on $k$ and $\tau$ alone gives us a runtime of $g(k,\tau)\log^{O(1)} r$ as required.
\end{proof}

We now give the proof of Theorem~\ref{thm:iqp}.

\begin{proof}
    We solve \textsc{SingMinReachDelete} as follows. For the input temporal graph $\mathcal{G}$, we find the temporal neighbourhood partition graph. This can be found in time $O(\Lambda n^3)$ by Proposition \ref{prop:poly-tnd}. We make a subclass for each time a class in the partition could be reached from the source, with an additional subclass for vertices in this class which are not temporally reachable from the source following a deletion; by Lemma \ref{cor:iqp-number-subclasses}, there are at most $k(k+1)\tau$ subclasses.
    For each possible assignment $\phi$ of subclasses to empty/non-empty, we find the sets of time-edges between two subclasses which must be deleted by Lemma \ref{lem:iqp-possible-phi}; there are at most $2^{k(k+1)\tau}$ possible functions $\phi$ to consider. Given the sets of time-edges we must delete and a function $\phi$, we then apply Lemma \ref{lem:iqp-cardinality-min} to find the number of vertices in each subset such that the number of vertices reached from the source is at most $r$ and the number of time-edges deleted is minimised. 
\end{proof}

\section{Conclusion and Open Questions}\label{sec:conclusion}
We have described three temporal parameters that form a hierarchy mirroring the one formed by their static analogues, and that can all be small when the temporal graph is dense at every timestep. We provide examples of problems demonstrating that there is a separation between the classes of problems admitting efficient algorithms when each of the parameters is bounded. 
As is the case for the corresponding static parameters, we expect that there will be many problems for which these temporal parameters give fixed-parameter tractability, and suggest exploration of temporal extensions of static problems for which there are known fpt-algorithms as future work.  From a practical perspective, it would also be interesting to investigate the values of these new parameters on dense real-world temporal networks.

One of the most celebrated results involving static cliquewidth is a metatheorem due to Courcelle et al. \cite{courcelle_linear_2000} which guarantees the existence of a linear-time algorithm for any problem expressible in a suitable fragment of logic (MSO$_1$) on graphs of bounded cliquewidth.
It is a natural question whether an analogous metatheorem exists for temporal cliquewidth.  
A promising approach for less restrictive logic might be to encode a temporal graph as an arbitrary relational structure (as has been done for a temporal version of treewidth \cite{fluschnik_as_2020}).  
A major challenge here, however, is that to the best of our knowledge there is no single notion of cliquewidth for relational structures: several alternatives have been introduced \cite{DBLP:journals/corr/abs-0806-0103,courcelle_graph_2012}, but none has all of the desirable properties.  Moreover, we believe that any encoding of a temporal graph of bounded temporal cliquewidth as a relational structure that preserves all the information in the original is unlikely to have bounded width for any cliquewidth-style measure unless we also bound the lifetime of the temporal graph. Nevertheless, this general direction merits further investigation, and there is potential for a useful metatheorem even if it is necessary to further restrict the fragment of logic considered or the structure of the temporal graph.

\section{Acknowledgements}
Samuel D. Hand is supported by an EPSRC doctoral training account. Kitty Meeks is supported by EPSRC grants  EP/T004878/1 and EP/V032305/1. Jessica Enright is supported by EPSRC grant EP/T004878/1.
   
   \bibliography{references}

\end{document}